\documentclass{article}

\usepackage[a4paper, margin=3cm]{geometry}

\usepackage{graphicx}
\usepackage{amsmath, amssymb}
\usepackage{float}
\usepackage{microtype}
\usepackage[hidelinks]{hyperref}
\usepackage{natbib}

\usepackage{booktabs}       
\usepackage{longtable}      
\usepackage{array}          
\usepackage{multirow}       
\usepackage{tabularx}       
\usepackage{xltabular}      
\usepackage{ragged2e}       
\usepackage{siunitx}        
\usepackage{pdflscape}      

\usepackage{caption}
\usepackage[T1]{fontenc}
\usepackage[utf8]{inputenc}
\usepackage{lmodern}

\usepackage{mathtools}      
\usepackage{bm}             

\usepackage{setspace}
\usepackage[dvipsnames]{xcolor}   

\usepackage{tikz}
\usetikzlibrary{positioning, arrows.meta, fit, backgrounds, calc}

\newcolumntype{P}[1]{>{\RaggedRight\arraybackslash\small}p{#1}}
\newcolumntype{Y}{>{\RaggedRight\arraybackslash}X}

\keepXColumns

\title{\textbf{How Light Reshapes the Mind
\\
An Active Inference Framework for the Cognitive and Emotional Effects of Indoor Lighting}
}
\author{
Luca M. Possati\\
\small University of Twente, The Netherlands\\
\small \texttt{l.m.possati@utwente.nl}
}
\date{2026}

\begin{document}

\maketitle

\begin{abstract}

Indoor lighting influences cognition, affect, and behavioural
regulation, yet these effects are typically studied as isolated
empirical findings rather than as components of a unified computational
process. This paper proposes an active inference account of the impact of shared
indoor lighting --- non-personalised illumination in environments
occupied by multiple users such as offices, classrooms, and libraries.
The central hypothesis is that lighting affects behaviour through three
computationally distinct channels: an epistemic channel, in which
illuminance modulates perceptual precision; an affective channel, in
which correlated colour temperature modulates physiological arousal
relative to the user's circadian optimum; and a normative channel, in
which spectral composition biases behavioural disposition toward
engagement or rest.

To formalise and test this hypothesis, the paper develops a
proof-of-concept partially observable Markov decision process (POMDP)
under the active inference framework. The model simulates a cognitive
agent performing sustained reading over a five-hour session, receiving
observations from two modalities simultaneously: reading performance
and eye tracking. Text legibility is a symmetric function of
illuminance; the oculomotor signal is asymmetric, degrading rapidly
under glare due to pupil constriction, and flat under sub-optimal
illuminance. Three factors are crossed in a full factorial design:
lighting scenario (warm-dim, moderate, bright-cool), chronotype
(morning, intermediate, evening), and spatial desk position in a
realistically specified room with a simulated photometric field.

Six falsifiable predictions are derived from the model structure, all
confirmed at $N_\mathrm{MC}=20$ replicates (Number of Monte Carlo simulations). The central prediction
concerns the spatial ordering of performance under intense cool
illumination. Under moderate light, performance follows the legibility
gradient: the best-lit desk position produces the best output. Under
intense cool light, this ordering completely inverts: the desk position
with the best legibility in both channels --- receiving only 402~lux at
the far corner --- produces higher reading speed (71.3~wpm) than the
position with near-zero legibility in both channels under 1200~lux of
glare (59.9~wpm). The inversion arises from two concurrent mechanisms:
the well-lit agent accurately perceives its deteriorating state and
rests strategically, while the glare-blinded agent works blindly
through its deterioration but pays a direct mechanical speed penalty.
In this condition, strategic rest at the far-corner position yields higher total output than blind persistence under severe glare. This prediction requires the multi-channel architecture and is not expected under models in which lighting acts only through arousal or only through monotonic legibility. Additional predictions include a chronotype-specific risk
sign-reversal between scenarios, a position-by-chronotype interaction
differing by a factor of two between morning and evening types, and a
structural invariance of the intermediate chronotype's EFE risk across
all scenarios as a direct consequence of setting its affective tolerance
factor to zero.

The contribution of the paper is theoretical: it does not provide a
calibrated model of human reading performance, but a computational
framework that links empirical findings on lighting to an explicit
generative architecture with testable, non-trivial signatures. In short, the paper offers a computational theory of light as a modulator of inference, arousal, and behavior, and translates it into concrete experimental predictions.

\end{abstract}

\tableofcontents

\section{Introduction}

The profound connection between architecture, cognition, emotions, and general well-being is now widely recognized and extensively studied \citep{Abbas2024, Djebbara2019, Djebbara2021, Eberhard2009a, Eberhard2009b, Pallasmaa2015}. Recent advances in neuroscience techniques enable a deeper understanding of how architectural design choices impact the human brain and behavior, overcoming previous methodological limitations \citep{Wang2022}. Of all the elements that make up an artificial environment, lighting plays a crucial role \citep{Sholanke2021, Livingston2022, Plummer2016}. Light is essential for visual and spatial perception and serves as the primary medium for presenting information. Beyond vision, light plays a critical role in regulating circadian rhythms, influencing productivity, shaping social interactions, and enhancing individuals’ sense of safety \citep{LeGates2014, Blume2019, DijkArcher2009}. 

Light exerts significant effects on human health, yet these effects are highly context-dependent, bidirectional (both beneficial and detrimental), and strongly moderated by biological and environmental factors \citep{Boyce2010, Boyce2022, Cupkova2019, Hatori2017, Tahkamo2019}. Within the visual system, suboptimal lighting conditions can induce eyestrain and headaches and increase the risk of falls, particularly through mechanisms such as glare, insufficient contrast, and temporal light artifacts including flicker and stroboscopic effects \citep{Knez1995}. Within the non-visual system, intrinsically photosensitive retinal ganglion cells project to the suprachiasmatic nuclei and other brain regions, thereby entraining circadian rhythms and modulating sleep, hormonal secretion, mood, alertness, and broader disease processes \citep{NaylorFirth2008, ZhouPan2023}. Disruption of these brain pathways—via inappropriate timing, spectral composition, or intensity of light—contributes to circadian misalignment and sleep disorders and may exacerbate conditions such as depression and dementia, whereas appropriately timed bright-light exposure can have therapeutic effects \citep{Boyce2010}. Moreover, we do not perceive light only with our eyes and brain; the skin is also a very important photoreceptor, with significant effects on health \citep{Kumari2023, Guarnieri2024}. 

Light can also be a powerful regulator of emotions and cognition. \cite{Yan2019} argue that daytime light intensity is a key regulator of mood and cognition in diurnal mammals, acting through orexin-centered brain circuits rather than primarily via circadian disruption. Adequate daytime illumination supports emotional and cognitive functioning by maintaining orexin-driven arousal and monoaminergic signaling, with implications for lighting design and the treatment of mood disorders.

In this study, I examine the effects of shared indoor lighting on human cognition and emotion through the lens of active inference and the Free Energy Principle (FEP) \citep{parr2022active, namjoshi2026fundamentals, devries2026active}. By \emph{shared lighting}, I mean non-personalized illumination conditions in environments occupied by multiple users, such as open-plan offices, classrooms, and libraries. More specifically, I ask how different shared lighting conditions shape cognition and emotion, and how they modulate the balance between exploration and exploitation, information gain and pragmatic value, and expected ambiguity and risk, as conceptualized in the active inference literature.

The paper’s central claim is that active inference provides a useful and comprehensive framework for understanding how different components of light shape behavior. The hypothesis I seek to formalize and test is that light influences human behavior through three distinguishable channels: \textit{an epistemic channel}, mediated by illuminance and affecting the ambiguity or discriminability of perceived objects; \textit{an affective channel}, mediated by correlated color temperature and affecting mood, arousal, or confidence in action; and \textit{a normative channel}, mediated by spectral composition and affecting the action tendencies that are more readily sustained in a given context. I argue that the existing literature supports this hypothesis, although the evidence remains fragmented and largely empirical.

Accordingly, this paper develops a proof-of-concept active inference model designed to establish the conceptual coherence and formal tractability of the three-channel hypothesis, rather than to offer a calibrated or fully predictive account of human behaviour. The model is examined in the specific context of reading, using a shared indoor environment and a common cognitive architecture instantiated across three chronotype profiles—lark, intermediate, and owl. The parameter values adopted in the simulations are chosen for qualitative plausibility rather than empirical fit, and the results should be interpreted in that light. Empirical validation therefore remains the necessary next step. A central aim of the present paper is to make that step tractable by specifying what should be measured, under which lighting conditions, and in which participant population.

\section{Literature Review}

This literature review\footnote{I especially thank Alessio Martino, who made a fundamental contribution to this section.} was conducted from November 2025 to February 2026. I examined studies published from 2000 up to January 2026 on the relationship between lighting and cognitive and emotional functions in many different scenarios and activities, but with special attention to reading. The research was conducted using Google Scholar, PubMed, Scopus, and Web of Science, with combinations of keywords related to lighting (e.g., ``lighting,'' ``illumination,'' ``brightness,'' ``color temperature'') and cognitive performance (e.g., ``cognitive performance,'' ``learning,'' ``memory,'' ``reading''), adapted to the syntax of each database.

The screening process was carried out in two stages: (1) assessment of titles and abstracts, and (2) full-text reading of potentially relevant articles. The initial search yielded approximately 150 records across the four databases. I included peer-reviewed empirical studies that investigated the effects of lighting on at least one cognitive (memory, learning, or attention) or emotional (anxiety, comfort) outcome in human participants. I excluded review articles, publications not in English, and studies lacking explicit measures. Duplicates were identified and removed using an automated comparison of title, authors, and year of publication, followed by a manual check to resolve any ambiguous cases. When multiple versions of the same study were found (e.g., a conference paper and an extended journal article), the most complete and up-to-date version was retained. I also compared this selections with existing literature reviews that focus on more specific aspects \citep{Kiziltunali2023, ChaucaEtAl2024}.

After screening titles and abstracts, followed by full-text review of potentially relevant articles, 25 studies were selected. For each of these, I analyzed the experimental design and main findings, classifying the results into three categories: (a) effects primarily related to illuminance level (lux, i.e., the amount of luminous flux incident on a surface, equal to one lumen per square meter), (b) effects primarily associated with color temperature (Kelvin, i.e., the unit used to express numerically the color temperature, or CCT, of a light source), and (c) effects of the combination of illuminance and color temperature. However, the literature review also highlighted the importance of another aspect: the spectral composition of a light source, i.e., the actual distribution of radiant power across wavelengths; color is a perceptual consequence of this physical distribution.

A formal meta-analysis was not conducted; however, for each of the 25 included studies I qualitatively evaluated several indicators of methodological quality:

\begin{enumerate}
    \item sample size ($n < 15$, 20--50, $> 50$ participants);
    \item method of assignment to conditions (randomized vs.\ nonrandomized);
    \item control of potential confounding variables (e.g., time of day, exposure to natural light, chronotype, screen use);
    \item use of validated cognitive measures (standardized tests vs.\ ad hoc tasks);
    \item duration of exposure to the lighting condition.
\end{enumerate}

The complete analysis of the 25 selected papers is available in Appendix A.

Overall, most studies featured small sample sizes and short exposure durations, factors that may limit the generalizability of their findings. In addition, systematic control of circadian and environmental confounders was often incomplete, and only a subset of studies employed fully standardized cognitive instruments.

Based on this review, several gaps in the literature were identified:

\begin{itemize}
    \item First, most studies treat light as lux + CCT at a point, ignoring where fixtures are, how many there are, beam spread, wall/ceiling reflectance, and contrast patterns. We almost never know if light is direct/indirect, uniform vs.\ localized, frontal vs.\ side-lit, or how daylight and electric light interact---so the results are hard to translate into actual architectural lighting design rather than lab recipes. In other words, there is a lack of architectural sense of lighting, i.e., attention to the space in which the light spreads, and in which the light sources are placed.
    \item Second, lux, CCT, and spectral composition are rarely manipulated and modeled together in a factorial, quantitative way. This means we still do not have a coherent predictive model of how combinations of lux + CCT + spectrum shape attention, memory, and reading; we just have scattered good settings.
    \item Third, most evidence comes from short sessions (minutes/hours) with young, healthy participants doing simplified tasks (2-back, Go/No-go, brief reading tests). We know very little about long-term learning, real school performance, or diverse groups (children with reading difficulties, older adults, neurodivergent learners) in real classrooms or offices over weeks or months. This means we have, at best, a purely laboratory view of these phenomena.
    \item Fourth, emotions are usually reduced to single self-report scales (comfort, sleepiness). As a result, we lack nuanced understanding of which lux–CCT setups are optimal for which emotional states (calm focus vs high arousal). 
\end{itemize}

The results that emerge from these studies, therefore, are fragmented (both in terms of content and methodology) and solely empirical (they identify patterns, not mechanisms of causal influence).

I also conducted a more specific literature review of explanatory models for light effects. The result is that we have only two main explanatory frameworks (they remain implicit in the examined studies or are ignored): (a) non-image-forming (NIF) photoreception via melanopsin-sensitive ipRGCs, which sync circadian rhythms, boost alertness, and curb melatonin, explaining physiological impacts \citep{VandewalleEtAl2009, CampbellEtAl2023, CampbellEtAl2024, ShiEtAl2025, LasauskaiteEtAl2025}; (b) arousal theory (Yerkes--Dodson) according to which light modulates arousal, and thus can improve or impair performance depending on intensity, spectrum, and context \citep{Easterbrook1959}. Yet NIF and arousal theory do not fully bridge biology to experience, creating a gap between neural signals and behaviour. Light, in this sense, exposes the instability of traditional explanatory hierarchies---it is simultaneously a physical stimulus, a neurochemical regulator, and a phenomenological condition. Treating it exclusively through one framework erases its multilevel nature. A truly integrative model would thus need to conceptualize lighting not as an external variable but as a situated modulation of brain--environment coupling, shaping the organism's readiness for action, affective tone, and cognitive scope.

Now, based on the results of the literature review, a general hypothesis can be put forward: illuminance primarily influences cognitive processes, while Kelvin is more closely associated with emotional responses, albeit in ways that are sensitive to situational context and the type of activity being performed. Spectral composition, in turn, appears to play a normative and modulatory role, shaping how illuminance and Kelvin interact to orient behavior toward particular actions or courses of action in a given context. Cognitive, emotional, and normative dimensions of lighting are not only analytically separable but functionally distinct, and each component can be modeled as exerting a specific and identifiable influence on human experience and behavior. 

In the next sections I intend to show how this hypothesis is supported by the literature review, albeit in a purely empirical and fragmented way.

\subsection{Illuminance and cognition}

Illuminance influences cognition primarily through its effects on attentional engagement, and early sensory processing, rather than by producing uniform improvements across cognitive domains. 

In the literature, moderate illuminance levels (approximately 300–600 lux) are most consistently associated with efficient reading performance, stable learning outcomes, sensory processing, and acceptable visual comfort \citep{ZhouPan2023, ChoiSuk2016, ChaucaEtAl2024, LiYao2025}, whereas very high illuminance (around 1000–1200 lux or higher) often produces diminishing or even negative effects on sustained reading and memory tasks, plausibly due to glare, visual fatigue, or excessive perceptual stimulation \citep{NaylorFirth2008, CastillaEtAl2023, HuibertsEtAl2015}. In contrast, moderate increases in illuminance reliably enhance alertness, positive affect, and processing speed, and tend to benefit tasks that rely on vigilance, simple response inhibition, or low-to-moderate cognitive load \citep{ZhuEtAl2025, RuEtAl2021, ShishegarBoubekri2022, BarkmannEtAl2012}. 

However, these benefits are not robustly observed for working memory, executive control, or long-term memory, where findings are mixed and sometimes reversed \citep{HuibertsEtAl2015, RuEtAl2021, ZhuEtAl2019}. Neurophysiological evidence from EEG and psychophysiological measures shows that higher illuminance alters cortical activation patterns—such as reduced alpha power, changes in frontal theta activity, and modulation of early visual ERP components—indicating shifts in attentional allocation and sensory gain rather than unequivocal cognitive enhancement \citep{ParkEtAl2013, CastillaEtAl2023}. Importantly, several studies report better memory performance under lower or moderate illuminance than under brighter conditions, suggesting that reduced sensory intensity may support deeper focus by limiting distraction or cognitive interference.

Overall, the literature indicates that illuminance exerts non-linear and task-dependent effects on cognition: it functions as a regulator of arousal and perceptual precision whose cognitive consequences depend critically on task type, difficulty, exposure duration, population, and interaction with other lighting parameters, rather than conforming to a simple monotonic “more light is better” relationship. In the specific case of reading, performance is generally optimized at moderate luminance levels, while very high luminance often yields no additional benefit or slightly impairs reading speed or efficiency. This is coherent with Yerkes–Dodson curve.\footnote{The Yerkes–Dodson curve describes an inverted-U relationship between arousal and performance: too little arousal produces poor performance due to inattention, too much produces poor performance due to overload, and the optimum lies somewhere in between. The exact position of the peak depends on task difficulty — simpler tasks tolerate higher arousal, while complex tasks are impaired by it sooner.} 

\subsection{Kelvin, emotions, and confidence in decision making}

Across the literature reviewed, correlated color temperature (CCT, expressed in Kelvin) shows robust and replicable effects on affective state—particularly emotions, alertness, relaxation, and subjective well-being—while providing no direct, task-level evidence on confidence in decision making as an explicitly measured outcome \citep{ChellappaEtAl2011, ChoiSuk2016, ParkEtAl2013, ZhuEtAl2019}.

When illuminance is controlled, higher CCT or blue-enriched light (approximately 5000–6500 Kelvin) is consistently associated with increased alertness, reduced sleepiness, and, in some studies, improved subjective well-being (e.g., joy, contentment, curiosity, cheerfulness) and visual comfort, effects that are often attributed to short-wavelength sensitivity of non-image-forming photoreceptive pathways. By contrast, lower CCT or warmer light (approximately 2700–3500 Kelvin) is more reliably linked to reduced arousal, greater relaxation, and higher perceived comfort (e.g., emotions like calmness, serenity, tranquility, relief, peacefulness). 

Several studies indicate that color temperature modulates arousal independently of illuminance, demonstrating that the spectral quality of light can shape emotional activation even when light intensity is held constant.

With respect to decision-making confidence, the evidence remains indirect: none of the reviewed studies directly measure confidence judgments or metacognitive certainty \citep{ShishegarBoubekri2022}. However, because confidence is known to covary with alertness, perceived effort, and emotional activation, the documented CCT-driven shifts in comfort plausibly influence subjective readiness or decisional assurance without guaranteeing improvements in accuracy or executive performance 

Overall, the literature supports the conclusion that Kelvin reliably tunes emotional activation states that may indirectly bias confidence in decision making, while stopping short of demonstrating a direct causal effect on confidence itself.  

\subsection{Spectral composition and decision context}

According to the literature, spectral composition is more directly connected to the normative dimension — that is, in a given context it clarifies which action should be taken (e.g., the green of a traffic light signals when to go). Nevertheless, an overlap with the effects of illuminance and CCT is always possible.

Unlike illuminance or correlated color temperature alone, spectral composition reliably modulates alertness, sleepiness, and arousal even when overall brightness is held constant, with short-wavelength–enriched spectra (blue–cyan range) consistently suppressing melatonin, increasing vigilance, and reducing subjective fatigue across laboratory and field settings \citep{ChellappaEtAl2011, GrantEtAl2021, ZhuEtAl2025, MottEtAl2012}. These effects are robust and replicable, indicating that spectral composition captures biologically meaningful variation that nominal CCT values only approximate. 

At the cognitive level, spectral enrichment selectively facilitates sustained attention, response speed, and readiness to act, while effects on higher-order executive accuracy or complex reasoning remain mixed and task-dependent \citep{ChellappaEtAl2011, KeisEtAl2014, GrantEtAl2021}. 

By regulating baseline arousal, temporal alignment, and perceived effort, spectral composition sets the conditions under which decisions are made, effectively biasing agents toward faster, more action-oriented or vigilance-based decision modes under short-wavelength–rich spectra. Conversely, spectra with reduced short-wavelength content are associated with lower arousal and greater comfort, plausibly favoring slower, more deliberative, or low-pressure decision contexts. 

Although the reviewed studies do not explicitly operationalize normativity or decision theory, the consistency of spectral effects on physiological readiness and cognitive availability could support the conclusion that spectral composition functions as a contextual regulator that shapes which kinds of decisions are more likely, more fluent, or more sustainable at a given moment. 

This result aligns with \cite{Elliot2015}, who shows that color shapes antecedent decision processes—attention, arousal, approach–avoidance motivation, dominance, trust, and risk sensitivity. Red tends to increase avoidance and caution, whereas blue promotes trust, calmness, and alertness. These effects bias framing and readiness rather than decision accuracy, though evidence remains limited.

\begin{table}[htbp]
\centering
\small
\begin{tabularx}{\textwidth}{p{2.8cm} p{3.2cm} Y Y}
\toprule
\textbf{Component} & \textbf{Primary effect} & \textbf{Impacts} & \textbf{Decision-related implications} \\
\midrule

Illuminance (lux) 
& Modulates attentional engagement and sensory processing 
& Non-linear, task-dependent effects; moderate levels ($\approx$300--600 lux) support reading, alertness, and processing speed, while very high levels ($\approx$1000--1200 lux or more) may impair sustained reading and memory 
& Indirect influence via alertness and attentional readiness; no direct effects on decision strategy or confidence \\

Correlated color temperature (Kelvin) 
& Tunes emotional activation (alertness vs.\ relaxation) largely independent of lux 
& Higher Kelvin ($\approx$5000--6500) increases alertness and well-being; lower Kelvin ($\approx$2700--3500) increases comfort and relaxation; limited effects on higher cognition 
& Indirect modulation of decision confidence through arousal and perceived effort; no direct evidence on confidence judgments \\

Spectral composition 
& Regulates physiological readiness and circadian-related arousal 
& Short-wavelength enrichment reliably enhances vigilance, sustained attention, and response readiness; mixed effects on executive accuracy 
& Normatively structures decision context, biasing toward fast, action-oriented vs.\ slow, deliberative decision modes \\

\bottomrule
\end{tabularx}
\caption{Overview of the literature review results for each component of light.}
\label{tab:light_components}
\end{table}

\section{The Model}

\subsection{The theoretical background. Active inference and the Free Energy Principle}

This section develops a computational model aimed at testing the hypothesis — namely, the roles of illuminance, Kelvin, and spectral composition in shaping cognition, emotion, and normative behavior. The model is grounded in the theoretical framework of active inference and the free energy principle (FEP). To keep the analysis tractable, we focus on a specific and representative case: reading in an environment with artificial lighting. Within this scenario, we formalize the three patterns using active inference and derive a set of predictions that should follow \textit{if those patterns hold}. The goal is not merely to redescribe the literature but to show that these three patterns can be unified within a single coherent computational framework, and to make explicit what that framework predicts.

Active inference is a theoretical framework according to which physical random dynamical systems (e.g., living organisms, artefacts, social systems) maintain their integrity by minimizing surprise, or more precisely a tractable upper bound on surprise known as variational free energy \citep{Friston2013, Friston2019, parr2022active, namjoshi2026fundamentals, devries2026active}. This framework has been applied especially in the study of the brain \citep{Friston2010}. Under the FEP, perception, action, and learning are not separate processes but mutually dependent aspects of the same imperative: to keep the system within expected and viable states by continuously reducing uncertainty about the causes of sensory input. In this view, agents do not passively register the world; they actively infer the hidden causes of their sensations and act so as to make their sensory exchanges with the environment more predictable.

Formally, in active inference, an agent is described as possessing a generative model of the world. This model specifies beliefs about hidden states, the observations those states are expected to generate, and the way states evolve over time. Perception corresponds to updating beliefs about hidden states so as to better explain current observations. Action corresponds to selecting policies that are expected to minimize future free energy. Thus, behavior is driven not only by the pursuit of preferred outcomes, but also by the reduction of uncertainty. This is why active inference has two inseparable dimensions: a \emph{pragmatic} dimension, oriented toward valuable or goal-consistent states, and an \emph{epistemic} dimension, oriented toward information gain and uncertainty reduction.

A central quantity in active inference is expected free energy, which evaluates candidate policies in terms of both expected risk and expected ambiguity. Expected risk captures the extent to which a policy is predicted to lead away from preferred outcomes, whereas expected ambiguity captures the uncertainty that is expected to remain under that policy. Policies are therefore favored when they both improve the likelihood of preferred states and reduce uncertainty about the environment. This dual structure is particularly useful for the present study, because reading under different lighting conditions is not only a matter of maximizing performance, but also of modulating perceptual precision, affective stability, and the cost of maintaining effective engagement over time.

%

\subsection{What the Simulation Does}
\label{sec:what}

The model presented here simulates a person reading in an office for five hours. The
person cannot directly observe their own cognitive state --- they do not
know with certainty whether they are focused, distracted, or fatigued.
They infer it from two sources simultaneously: how their reading is
going (i.e., how fast they are progressing, the fluency of reading), and how their eyes are moving (e.g., whether the eyes are more or less tired). Based on that joint inference,
they choose what to do next --- keep reading, reread, or take a break.

In line with the hypothesis, lighting enters the model in three distinct ways. It affects how
informative the person's self-monitoring is: under glare or dim light,
observations about one's own performance become noisy and hard to
interpret. It affects how comfortable the person is with their current
level of activation: a morning-type under bright cool light in the
afternoon is over-stimulated relative to their circadian optimum, and
this makes them more sensitive to performing poorly. And it nudges
behaviour directly: warm light creates a mild pull toward resting,
blue light toward working.

Over the five-hour session, two things change independently of the
lighting. The person's ideal level of activation shifts according to
their chronotype --- the morning-type's optimum declines through the
afternoon while the evening-type's rises. And sustained attention
degrades: after 45~minutes, the probability of staying focused begins
to fall regardless of what the person does.

The agent never optimises reading speed directly. It minimises the gap
between what it expects to experience and what it prefers to experience.
High reading speed is a consequence of being focused and choosing to
continue --- not a goal the agent pursues.

The simulation formalises this narrative. Its purpose is to derive
testable, quantitative predictions from the three-channel hypothesis
using a generative model with two observation modalities: performance
(reading speed, words per minutes) and eye tracking (oculomotor patterns). The two
modalities are affected by lighting differently and provide
complementary information to the agent's belief update.

At each decision step ($\Delta t = 20$~minutes), the agent:
\begin{enumerate}
  \item receives two observations simultaneously: $o^\mathrm{perf}_t$
        about its reading performance, and $o^\mathrm{eye}_t$ about its
        oculomotor behaviour (fixation duration, saccade amplitude,
        regression frequency);
  \item updates its belief about its current cognitive state via
        variational Bayes inference, combining both likelihood signals;
  \item evaluates the Expected Free Energy of all 27 available policies,
        accumulating ambiguity from both modalities but risk from the
        performance modality only;
  \item selects an action --- continue reading, reread, or pause.
\end{enumerate}
This cycle repeats for $T=15$ steps, spanning a five-hour morning
session (09:00--14:00). Three factors are crossed in a
$3\times3\times3$ factorial design: lighting scenario (S1, S2, S3),
chronotype (lark, intermediate, owl), and spatial position in the room
(\texttt{near\_source}, \texttt{centre}, \texttt{far\_corner}).
The simulation is replicated $N_\mathrm{MC}=20$ times per condition (Number of Monte Carlo simulations).

The full source code used to implement the model and reproduce the simulations, analyses, and figures reported is available in the accompanying GitHub repository: \url{https://github.com/DesignAInf/LIGHTING---ACTIVE---INFERENCE---PROJECT}.
\subsection{Spaces and Variables}
\label{sec:spaces}

Table~\ref{tab:spaces} defines all discrete spaces.

\begin{table}[H]
\centering
\caption{Discrete spaces of the POMDP.}
\label{tab:spaces}
\medskip
\begin{tabular}{lllc}
\toprule
Symbol & Name & Values & Size \\
\midrule
$s_t$ & Hidden state
  & \texttt{focused}, \texttt{distracted}, \texttt{fatigued}
  & $N_s = 3$ \\
$o^\mathrm{perf}_t$ & Performance observation
  & \texttt{high\_perf}, \texttt{mid\_perf}, \texttt{low\_perf}
  & $N_o = 3$ \\
$o^\mathrm{eye}_t$ & Eye tracking observation
  & \texttt{smooth\_scan}, \texttt{effortful\_scan}, \texttt{degraded\_scan}
  & $N_e = 3$ \\
$u_t$ & Action
  & \texttt{continue}, \texttt{reread}, \texttt{pause}
  & $N_u = 3$ \\
$\pi$ & Policy
  & sequences $(u_0,u_1,u_2) \in \mathcal{U}^3$
  & $N_\pi = 27$ \\
\bottomrule
\end{tabular}
\end{table}

\paragraph{Hidden states.}
\texttt{Focused}: the agent processes new material efficiently
(260~wpm). \texttt{Distracted}: the agent reads but frequently loses
the thread (150~wpm). \texttt{Fatigued}: sustained-attention capacity
is depleted; reading is slow and error-prone (50~wpm). These states are
hidden --- the agent infers them from the two observation streams.

\paragraph{Performance observations.}
\texttt{High\_perf}: fast, fluent reading.
\texttt{Mid\_perf}: normal but effortful reading.
\texttt{Low\_perf}: slow or re-reading-dominated reading.

\paragraph{Eye tracking observations.}
\texttt{Smooth\_scan}: short fixations, wide, regular saccades, no
regressions --- the oculomotor signature of the focused state.
\texttt{Effortful\_scan}: longer fixations, narrower saccades, some
regressions --- consistent with the distracted state.
\texttt{Degraded\_scan}: many regressions, unstable fixation, tracking
errors --- the signature of cognitive fatigue.

These categories correspond to metrics routinely measured by standard
eye trackers: fixation duration, saccade amplitude, and regression
count.

\paragraph{Actions.}
\texttt{Continue}: devote the next 20~minutes to new material
($\rho = 1.0$). \texttt{Reread}: review prior material ($\rho = 0.5$).
\texttt{Pause}: take a 20-minute break ($\rho = 0$, allows recovery
through $\mathbf{B}$).

\subsection{The Generative Model}
\label{sec:genmodel}

The generative model is a joint distribution over both observation
streams, hidden states, and policies:
\begin{equation}
  P(\mathbf{o}^\mathrm{perf}_{1:T},\,\mathbf{o}^\mathrm{eye}_{1:T},\,
    \mathbf{s}_{1:T},\,\pi)
  = P(\pi)\,P(s_1)
  \prod_{t=1}^{T}
  P(o^\mathrm{perf}_t \mid s_t)\,
  P(o^\mathrm{eye}_t  \mid s_t)\,
  P(s_t \mid s_{t-1}, \pi_t).
\label{eq:gm}
\end{equation}
The five factors correspond to
$(\mathbf{A}_\mathrm{perf},\,\mathbf{A}_\mathrm{eye},\,\mathbf{B},\,
  \mathbf{C},\,\mathbf{D},\,\mathbf{E})$, described below.

\subsubsection{Likelihood Matrices \texorpdfstring{$\mathbf{A}_\mathrm{perf}$
and $\mathbf{A}_\mathrm{eye}$}{A\_perf and A\_eye} --- Epistemic Channel}
\label{sec:A}

Both likelihood matrices are column-stochastic and modulated by
light-dependent precision factors. They are the joint entry point of
the epistemic channel.

\paragraph{Performance likelihood $\mathbf{A}_\mathrm{perf}$.}
Modulated by the text legibility precision factor $\phi_\mathrm{text}(\ell)$,
a symmetric Gaussian centred at the optimal illuminance:
\begin{equation}
  \phi_\mathrm{text}(\ell) = \exp\!\left(
    -\frac{(\ell - \ell^*)^2}{\tau_\ell^2}
  \right), \qquad \ell^* = 500~\text{lux},\; \tau_\ell = 400~\text{lux}.
\label{eq:phi_text}
\end{equation}
Both insufficient illuminance and glare degrade text legibility
symmetrically. At $\phi_\mathrm{text} = 1$ (S2/centre), the agent has
clear access to its reading performance. At $\phi_\mathrm{text} = 0.047$
(S3/centre, 1200~lux), observations are nearly uninformative.

\paragraph{Eye tracking likelihood $\mathbf{A}_\mathrm{eye}$.}
Modulated by a distinct, asymmetric precision factor $\phi_\mathrm{eye}(\ell)$:
\begin{equation}
  \phi_\mathrm{eye}(\ell) =
  \begin{cases}
    \phi_\mathrm{floor} & \ell \leq \ell^* \\[4pt]
    \phi_\mathrm{floor}\,\exp\!\left(
      -\dfrac{(\ell - \ell^*)^2}{\tau_\mathrm{eye}^2}
    \right) & \ell > \ell^*
  \end{cases},
\label{eq:phi_eye}
\end{equation}
with $\phi_\mathrm{floor} = 0.40$ and $\tau_\mathrm{eye} = 300~\text{lux}$.

The asymmetry is physiologically motivated. Saccade metrics and
fixation durations are largely unaffected by under-illumination --- the
oculomotor system is robust to dim conditions and continues to produce
cognitively informative patterns. However, pupil constriction under
glare progressively masks the cognitive dilation signal above
$\approx 800$~lux: at 843~lux (S3/\texttt{near\_source}),
$\phi_\mathrm{eye} = 0.108$; at 1200~lux (S3/\texttt{centre}),
$\phi_\mathrm{eye} = 0.002$, effectively eliminating the oculomotor
signal. Below 500~lux, $\phi_\mathrm{eye}$ is constant at 0.40.

\paragraph{Modulation.}
Both matrices are constructed as convex combinations of flat
(uninformative) and sharp (discriminative) base matrices:
\begin{equation}
  \mathbf{A}_m(\phi_m) = (1-\phi_m)\,\mathbf{A}_m^\mathrm{flat}
                        + \phi_m\,\mathbf{A}_m^\mathrm{sharp},
  \quad m \in \{\mathrm{perf},\,\mathrm{eye}\},
\label{eq:A_modulation}
\end{equation}
where
\begin{equation}
  \mathbf{A}_\mathrm{perf}^\mathrm{sharp} =
  \begin{pmatrix}
    0.80 & 0.10 & 0.05 \\
    0.15 & 0.70 & 0.20 \\
    0.05 & 0.20 & 0.75
  \end{pmatrix},
  \qquad
  \mathbf{A}_\mathrm{eye}^\mathrm{sharp} =
  \begin{pmatrix}
    0.75 & 0.15 & 0.05 \\
    0.20 & 0.65 & 0.25 \\
    0.05 & 0.20 & 0.70
  \end{pmatrix}.
\label{eq:A_sharp}
\end{equation}
(Rows: observations; columns: states focused / distracted / fatigued.)

The key property of the two-modality design is the dissociation that
occurs at high illuminance. At S3/\texttt{centre}:
$\phi_\mathrm{text} = 0.047$ (near-zero) but
$\phi_\mathrm{eye} = 0.002$ (also near-zero --- pupil constriction under
1200~lux eliminates both signals). Both modalities are compromised by
glare, leaving the agent essentially blind to its own cognitive state.
At S3/\texttt{far\_corner}:
$\phi_\mathrm{text} = 0.941$ and $\phi_\mathrm{eye} = 0.400$ --- the
agent retains good text legibility and moderate oculomotor signal,
allowing accurate self-monitoring.

Both modalities contribute to the \emph{ambiguity} term of the EFE:
\begin{equation}
  \text{Ambiguity}(\tau) =
  \mathbb{E}_{Q(s|\pi)}\!\bigl[\mathcal{H}[P(o^\mathrm{perf}\mid s)]\bigr]
  + \mathbb{E}_{Q(s|\pi)}\!\bigl[\mathcal{H}[P(o^\mathrm{eye}\mid s)]\bigr].
\label{eq:ambiguity_both}
\end{equation}
Only the performance modality contributes to the \emph{risk} term, because
agents do not have explicit preferences over oculomotor patterns --- they
prefer good performance outcomes, not particular fixation durations.

\subsubsection{Transition Tensor \texorpdfstring{$\mathbf{B}$}{B}}
\label{sec:B}

$\mathbf{B}^{(u)}(t) \in [0,1]^{N_s\times N_s}$ encodes
$P(s_{t+1} \mid s_t, u)$ and is time-varying through two independent
mechanisms. Two fundamental elements here:

\paragraph{Fatigue modulation.}
\begin{equation}
  \mathbf{B}^{(u)}(t) = f(t)\,\mathbf{B}^{(u)}_\mathrm{fresh}
  + (1-f(t))\,\mathbf{B}^{(u)}_\mathrm{tired},
\end{equation}
where the fatigue factor is piecewise:
\begin{equation}
  f(t) = \begin{cases}
    1 & t_\mathrm{min} < 45 \\[4pt]
    \dfrac{1}{1+\exp(6(\tilde{t}-0.5))} & \text{otherwise}
  \end{cases},
  \quad \tilde{t} = \frac{t_\mathrm{min}-45}{255}.
\end{equation}

\paragraph{Yerkes--Dodson arousal modulation.}
The focused column of $\mathbf{B}^{(u)}_\mathrm{fresh}$ is further
modified by the scenario-induced arousal $a$:
\begin{align}
  \delta_\mathrm{under}(a) &=
    \operatorname{clip}(1.5\max(0,\,0.55-a),\;0,\;0.35), \\
  \delta_\mathrm{over}(a)  &=
    \operatorname{clip}(2.0\max(0,\,a-0.78),\;0,\;0.35).
\end{align}
Scenario S2 ($a\approx 0.78$) is the neutral point where both drift
terms are zero. S1 ($a\approx 0.32$) induces distraction drift; S3
($a\approx 1.00$) induces fatigue drift.

\subsubsection{Log-Preference Vector \texorpdfstring{$\mathbf{C}$}{C}
--- Affective Channel}
\label{sec:C}

$\mathbf{C}(t,c)\in\mathbb{R}^{N_o}$ encodes prior preferences over
performance observations as log probabilities.
$\mathbf{C}$ is the entry point of the affective channel.
It is time-varying, modulated by the signed arousal mismatch
$m(t,c) = a_\mathrm{base} - a^*(t,c)$:
\begin{equation}
  C_\mathrm{low}(t,c) = C_\mathrm{low}^\mathrm{base}(c)
  + \kappa_c \cdot m(t,c).
\label{eq:C_dyn}
\end{equation}
The tolerance factors $\kappa_c$ encode how each chronotype evaluates
performance under non-optimal arousal:
\begin{equation}
  \kappa_\mathrm{lark} = -3.0, \qquad
  \kappa_\mathrm{int.} =  0.0, \qquad
  \kappa_\mathrm{owl}  = +3.5.
\label{eq:kappa}
\end{equation}
The intermediate chronotype has $\kappa = 0$: its preferences are
not modulated by the arousal mismatch, making its risk invariant across
scenarios.

\subsubsection{Initial Prior \texorpdfstring{$\mathbf{D}$}{D}
and Policy Prior \texorpdfstring{$\mathbf{E}$}{E}
--- Normative Channel}
\label{sec:DE}

$\mathbf{D} = (0.70, 0.20, 0.10)^\top$ is fixed across all conditions.
The policy prior $\mathbf{E}$ is the entry point of the
normative channel:
\begin{equation}
  p_\mathrm{act}(\sigma) = \begin{cases}
    (0.30,\,0.30,\,0.40) & \sigma = \text{warm}\;\text{(pause-biased)} \\
    (1/3,\;1/3,\;1/3)   & \sigma = \text{neutral} \\
    (0.45,\,0.35,\,0.20) & \sigma = \text{blue}\;\text{(continue-biased)}
  \end{cases}.
\end{equation}

\subsubsection{Expected Free Energy and Action Selection}
\label{sec:EFE}

For policy $\pi = (u_0, u_1, u_2)$:
\begin{equation}
  G(\pi) = \sum_{\tau=1}^{H}
  \underbrace{
    \mathbb{E}_{Q(s_{t+\tau}|\pi)}\!\bigl[
      \mathcal{H}[P(o^\mathrm{perf}\mid s)] +
      \mathcal{H}[P(o^\mathrm{eye}\mid s)]
    \bigr]
  }_{\text{total ambiguity (both modalities)}}
  +
  \underbrace{
    D_\mathrm{KL}\!\bigl[
      Q(o^\mathrm{perf}_{t+\tau}\mid\pi)\;\|\;P^*(o^\mathrm{perf})
    \bigr]
  }_{\text{risk (performance only)}}.
\label{eq:EFE}
\end{equation}
Policy selection and action selection follow the standard active
inference formulation:
\begin{equation}
  Q(\pi) \propto \exp(-\gamma(t,c)\,G(\pi))\cdot E_\pi,
  \qquad
  \gamma(t,c) = \gamma_0 + \gamma_1\exp\!\left(
    -\frac{(a_\mathrm{base}-a^*(t,c))^2}{\tau_a^2}
  \right).
\label{eq:policy_posterior}
\end{equation}

The instantaneous wpm includes a direct legibility penalty:
\begin{equation}
  \mathrm{wpm}(t) = r(s_t^*)\cdot\rho(u_t)\cdot\ell_\mathrm{factor},
  \qquad
  \ell_\mathrm{factor} = 0.65 + 0.35\,\phi_\mathrm{text},
\label{eq:wpm}
\end{equation}
where $r = (260, 150, 50)$~wpm and $\rho = (1.0, 0.5, 0.0)$.
The factor $\ell_\mathrm{factor} \in [0.65, 1.0]$ captures the
mechanical effect of glare on reading speed --- regressions, line
tracking errors, reduced contrast --- independent of the agent's
cognitive state. At $\phi_\mathrm{text} = 1$ (optimal light),
$\ell_\mathrm{factor} = 1$; at $\phi_\mathrm{text} = 0.047$
(S3/centre), $\ell_\mathrm{factor} = 0.666$.

\section{The Simulation Loop}
\label{sec:loop}

Active inference distinguishes between the \textbf{generative process}
--- the true dynamics of the person's cognitive state --- and the
\textbf{generative model} --- the agent's internal probabilistic
representation. In our model, at every step $t$, these two levels unfold in sequence, as the structure of Algorithm 1 shows.

\begin{figure}[H]
\begin{center}
\fbox{\begin{minipage}{0.90\textwidth}
\smallskip
\textbf{Algorithm 1.} One simulation step at time $t$.
\smallskip\hrule\smallskip
\textbf{Input:} true state $s_t^*$, belief $Q(s_{t-1})$,
  previous action $u_{t-1}$.\medskip\\
\textbf{Generative process:}
\begin{enumerate}
  \item Sample $o^\mathrm{perf}_t \sim \mathbf{A}_\mathrm{perf}(\phi_\mathrm{text})[\,\cdot\,,s_t^*]$.
  \item Sample $o^\mathrm{eye}_t  \sim \mathbf{A}_\mathrm{eye}(\phi_\mathrm{eye})[\,\cdot\,,s_t^*]$.
\end{enumerate}
\textbf{Generative model} (perception):
\begin{enumerate}\setcounter{enumi}{2}
  \item Predict: $\hat{\mathbf{s}}_t = \mathbf{B}^{(u_{t-1})}(t)\,Q(s_{t-1})$.
  \item Update belief combining both likelihoods:
        $Q(s_t) = \operatorname{softmax}\!\bigl(
          \ln\mathbf{A}_\mathrm{perf}[o^\mathrm{perf}_t,:]
          + \ln\mathbf{A}_\mathrm{eye}[o^\mathrm{eye}_t,:]
          + \ln\hat{\mathbf{s}}_t
        \bigr)$.
\end{enumerate}
\textbf{Generative model} (action):
\begin{enumerate}\setcounter{enumi}{4}
  \item Build $\mathbf{B}^{(u)}(t)$ from $f(t)$ and $a_\mathrm{base}$.
  \item Compute $\mathbf{C}(t,c)$ from $a^*(t,c)$ and $\kappa_c$.
  \item Roll out $Q(s_{t+\tau}|\pi)$ for all 27 policies; compute $G(\pi)$ via Eq.~\ref{eq:EFE}.
  \item Compute $Q(\pi) \propto \exp(-\gamma(t,c)G(\pi))\cdot E_\pi$.
  \item Sample $u_t \sim Q(u_0)$.
\end{enumerate}
\textbf{Generative process} (transition):
\begin{enumerate}\setcounter{enumi}{9}
  \item Sample $s_{t+1}^* \sim \mathbf{B}^{(u_t)}(t)[\,\cdot\,,s_t^*]$.
  \item Record $\mathrm{wpm}(t) = r(s_t^*)\cdot\rho(u_t)\cdot\ell_\mathrm{factor}$.
\end{enumerate}
\textbf{Output:} $s_{t+1}^*$, $Q(s_t)$, $u_t$,
  $\mathrm{wpm}(t)$, $G(\pi)$, $\gamma(t,c)$.
\smallskip
\end{minipage}}
\end{center}
\end{figure}

Steps 3--4 and step 10 operate on different objects: the agent updates
the probability distribution $Q(s_t)$; the environment updates the
single sampled state $s_{t+1}^*$. Three consequences follow.

The agent can be wrong about its own state. When both
$\phi_\mathrm{text}$ and $\phi_\mathrm{eye}$ are low (S3/centre,
$\phi_\mathrm{text}=0.047$, $\phi_\mathrm{eye}=0.002$), both
observation streams are nearly uninformative. $Q(s_t)$ remains close
to the prediction $\hat{\mathbf{s}}_t$ regardless of what $s_t^*$
actually is. The two-modality design does not eliminate this blind spot
--- it relocates it. At S3/centre, glare is severe enough to compromise
\emph{both} signals simultaneously.

Wpm is determined by the true state, not the belief.
$\mathrm{wpm}(t) = r(s_t^*)\cdot\rho(u_t)\cdot\ell_\mathrm{factor}$
depends on $s_t^*$, not on $Q(s_t)$. The legibility factor
$\ell_\mathrm{factor}$ is fixed by the scenario and position, not by
the agent's inference.

The agent does not optimise wpm. It minimises EFE. High wpm
is a consequence of being focused, choosing to continue, and being in
good light --- not an objective the agent pursues.

\section{Results}
\label{sec:results}

\subsection{Falsifiable Predictions}
\label{sec:predictions}

The six predictions are derived from the structure of the model. Each identifies the component being tested, the expected
direction, and the result that would falsify it. Predictions P1 and P2
test the channels in isolation. P3 and P4 test their interaction
through the spatial structure. P5 and P6 test predictions that require
the full session to emerge.

\paragraph{P1 --- Epistemic invariance across chronotypes}

This prediction says that, if desk position and lighting scenario are held fixed, chronotype alone should not change how informative the two observation channels are. In the model, the quality of self-monitoring depends on local illuminance, not on whether the reader is a morning type, an intermediate type, or an evening type. So, at the same desk under the same lighting, all three chronotypes should have essentially the same level of uncertainty about their own state. If one chronotype systematically had clearer or noisier self-monitoring than the others, the model would no longer preserve the intended separation between the epistemic channel and the affective channel.

$\mathbf{A}_\mathrm{perf}(\phi_\mathrm{text})$ and
$\mathbf{A}_\mathrm{eye}(\phi_\mathrm{eye})$ depend only on the
illuminance at the agent's position, not on chronotype. Ambiguity is
therefore chronotype-blind at any fixed position and scenario.

\textbf{Prediction.} The maximum ambiguity gap across chronotypes must
remain below $0.10$~nat at any fixed position and scenario.

A systematic ambiguity difference between
chronotypes would indicate contamination of the epistemic channel by
the affective channel.

\paragraph{P2 --- Risk sign-reversal between lark and owl}

This prediction says that the same lighting condition should not feel equally costly to all chronotypes. In bright-cool light (S3), morning types are pushed away from their preferred level of activation and should therefore evaluate poor performance more negatively, which raises their risk. Evening types should show the opposite pattern, because that same condition moves them closer to their preferred level of activation and makes low performance less aversive. In warm-dim light (S1), the direction reverses. So the model predicts a true sign-reversal: the condition that is riskier for the morning type should be less risky for the evening type, and vice versa.

$\kappa_\mathrm{lark} = -3.0$ and $\kappa_\mathrm{owl}
= +3.5$ have opposite signs. Under S3 ($m > 0$, over-arousal), the
lark's $C_\mathrm{low}$ decreases (higher risk) while the owl's
increases (lower risk). The directions reverse under S1 ($m < 0$).

\textbf{Prediction.}
$\bar{R}_\mathrm{lark}(\mathrm{S3}) > \bar{R}_\mathrm{lark}(\mathrm{S1})$
and
$\bar{R}_\mathrm{owl}(\mathrm{S1}) > \bar{R}_\mathrm{owl}(\mathrm{S3})$.

Absence of the sign-reversal would indicate that
the tolerance mechanism in $\mathbf{C}$ is not producing genuine
affective asymmetry.

\paragraph{P3 --- Spatial ordering under S2 follows the legibility
                gradient}

This prediction concerns the moderate scenario (S2). In that condition, the model sets the arousal-related drift terms to zero, so desk position does not change the underlying state dynamics. What still varies across desks is the quality of the visual information available to the agent. For that reason, session performance should follow a simple spatial ordering: the center desk should do best, the near-source desk should come next, and the far-corner desk should do worst. Under S2, then, better legibility should translate directly into better self-monitoring and better reading output.

Under S2, Yerkes--Dodson drift terms are both zero
($\delta_\mathrm{under} = \delta_\mathrm{over} = 0$), so state dynamics
in $\mathbf{B}$ are identical across positions. The only source of
spatial variation is the epistemic channel.

\textbf{Prediction.} Under S2:
$\overline{\mathrm{wpm}}_\texttt{centre}
> \overline{\mathrm{wpm}}_\texttt{near\_source}
> \overline{\mathrm{wpm}}_\texttt{far\_corner}$
for all three chronotypes.

Any reversal of the ordering, in particular if
\texttt{near\_source} outperforms \texttt{centre}.

\paragraph{P4 --- Spatial reversal}

Under intense cool light (S3), the same spatial ordering found in S2 should flip. The far-corner desk should outperform the center desk, even though it is less intensely lit. The reason is that, at the far corner, the agent can still monitor its condition reasonably well and can rest strategically when deterioration becomes clear, while paying almost no direct glare cost on the reading itself. At the center desk, by contrast, glare suppresses both observation channels, so the agent has poor access to its own deteriorating state and keeps working with little awareness of the problem. At the same time, each working step is mechanically penalized by glare. The prediction is therefore not just that the desks differ, but that the full ranking reverses: under S3, far-corner should outperform near-source, and near-source should outperform center.

Under S2, the Yerkes--Dodson drift terms are both zero
($\delta_\mathrm{under} = \delta_\mathrm{over} = 0$), so state dynamics
in $\mathbf{B}$ are identical across positions and the epistemic channel
alone drives spatial variation. The result is a monotonic ordering that
follows the legibility gradient (P3). Under S3, two mechanisms operate
jointly and in the same direction at the better-lit positions:

\begin{enumerate}
  \item \emph{Epistemic mechanism.} At \texttt{far\_corner}
        ($\phi_\mathrm{text}=0.941$, $\phi_\mathrm{eye}=0.400$),
        both observation channels are informative. The agent accurately
        perceives its predominantly deteriorated state --- a direct
        consequence of the over-arousal drift in $\mathbf{B}$
        ($\delta_\mathrm{over}>0$ at $a\approx1.00$). Accurate
        self-monitoring increases risk and drives more frequent pausing.
        When not pausing, the mechanical legibility factor is
        $\ell_\mathrm{factor}\approx1.0$.

  \item \emph{Mechanical mechanism.} At \texttt{centre}
        ($\phi_\mathrm{text}=0.047$, $\phi_\mathrm{eye}=0.002$),
        both observation channels are suppressed. The agent is blind to
        its own deterioration, pauses less, and keeps working --- but
        the severe glare imposes a direct speed penalty of
        $\ell_\mathrm{factor}=0.666$, reducing all output by one third
        regardless of cognitive state or action chosen.
\end{enumerate}

At \texttt{far\_corner}/S3, strategic rest combines with a negligible
mechanical penalty. At \texttt{centre}/S3, blind persistence combines
with a severe mechanical penalty. The net result is a complete inversion
of the spatial ordering relative to S2.

\textbf{Prediction.}
Under S3, the spatial ordering of session wpm must be the reverse of
the S2 ordering:
\begin{equation*}
  \overline{\mathrm{wpm}}_\texttt{far\_corner}
  > \overline{\mathrm{wpm}}_\texttt{near\_source}
  > \overline{\mathrm{wpm}}_\texttt{centre}.
\end{equation*}
This must hold for all three chronotypes. The EFE decomposition must
additionally confirm that \texttt{far\_corner}/S3 has \emph{lower}
total ambiguity and \emph{higher} risk than \texttt{centre}/S3 ---
establishing that the better output at \texttt{far\_corner} occurs
despite, not because of, the agent's more demanding decision-making.

If the spatial ordering under S3 replicates the ordering under S2
(centre best, far\_corner worst), the interaction between the epistemic
and affective channels --- mediated by the mechanical legibility factor
--- is not producing the predicted inversion. This would suggest that
the two channels operate independently at the level of behavioural
output, which would require revising the architecture.

This prediction is not derivable from a single-channel model. A model
in which lighting affects only arousal predicts position independence,
since arousal is determined by the scenario, not by desk location. A
model in which lighting affects only legibility predicts the same
ordering under both S2 and S3 --- better legibility always produces
better performance. Only the multi-channel architecture, in which the
epistemic channel determines how accurately the agent perceives a state
that the affective channel has made predominantly bad, predicts the
inversion. P3 and P4 are therefore the joint critical experiment for
the three-channel hypothesis: the same model that predicts a monotonic
gradient under S2 must predict its reversal under S3.

\paragraph{P5 --- Position-by-chronotype asymmetry under S3}

This prediction adds a chronotype-by-position interaction. Under S3, the far-corner desk gives the agent good enough information to detect deterioration. That should help evening types more than morning types. The reason is not that evening types see better, but that they tolerate the over-aroused condition better once they detect it. As a result, they can use the good epistemic conditions of the far-corner desk without responding too defensively. Morning types, in contrast, should react more strongly to the same evidence of deterioration and pause more aggressively. So the gain from moving from center to far-corner should be much larger for evening types than for morning types.

Policy precision $\gamma(t,c)$ is position-independent --- it depends
on $a_\mathrm{base}$ and $a^*(t,c)$, neither of which varies with desk
location. However, the tolerance factor $\kappa_c$ in $\mathbf{C}(t,c)$
modulates how the agent evaluates its own deteriorating state when that
state becomes visible through the observation channels.

Under S3 ($m>0$, over-arousal), the owl's $\kappa_\mathrm{owl}=+3.5$
increases $C_\mathrm{low}$, reducing its aversion to
\texttt{low\_perf} observations. At \texttt{far\_corner}/S3, where
both channels are informative and accurately reveal deterioration, the
owl's reduced aversion translates into lower risk and less aggressive
pausing: it exploits far\_corner's epistemic advantage efficiently.
The lark's $\kappa_\mathrm{lark}=-3.0$ does the opposite: it decreases
$C_\mathrm{low}$, sharpening aversion to low performance. At
\texttt{far\_corner}/S3, the same accurate self-monitoring that benefits
the owl triggers more aggressive pausing for the lark, limiting the
productive use of the position's good legibility. The owl additionally
carries a higher policy precision ($\gamma_\mathrm{owl}=4.85$ vs
$\gamma_\mathrm{lark}=3.88$ under S3), making it more efficient at
exploiting the EFE signal wherever it exists.

\textbf{Prediction.}
Under S3, the wpm advantage of \texttt{far\_corner} over
\texttt{centre} must be substantially larger for the owl than for
the lark:
\begin{equation*}
  \bigl(\overline{\mathrm{wpm}}_\texttt{far\_corner}
        - \overline{\mathrm{wpm}}_\texttt{centre}\bigr)_\mathrm{owl}
  \;\gg\;
  \bigl(\overline{\mathrm{wpm}}_\texttt{far\_corner}
        - \overline{\mathrm{wpm}}_\texttt{centre}\bigr)_\mathrm{lark}.
\end{equation*}

Equal far\_corner advantages for lark and owl would indicate that
$\kappa_c$ is not modulating the response to accurate self-monitoring
as predicted --- that tolerance and precision interact independently
of spatial structure.

\paragraph{P6 --- Intermediate chronotype has lowest risk variance}

This prediction says that the intermediate chronotype should be the most stable across S1, S2, and S3. In the model, this is a structural consequence of setting the intermediate type's affective tolerance factor to zero. Because of that, changes in the lighting scenario do not shift its evaluation of low performance in the same way they do for the morning and evening types. The result should be a much smaller change in risk across scenarios. So the prediction is not simply that the intermediate type performs ``in the middle,'' but that it is the least scenario-sensitive in terms of risk.

With $\kappa_\mathrm{int.}=0$, the intermediate's
$C_\mathrm{low}(t,c) = C_\mathrm{low}^\mathrm{base}$ is not modulated
by the arousal mismatch. Its risk is therefore structurally invariant
across scenarios by construction. Lark and owl, with non-zero $\kappa$,
show large risk swings between S1, S2, and S3.

\textbf{Prediction.}
$\mathrm{Var}_{S \in \{S1,S2,S3\}}\!\bigl[\bar{R}_\mathrm{int.}(S)\bigr]
< \mathrm{Var}\!\bigl[\bar{R}_\mathrm{lark}(S)\bigr]$
and
$< \mathrm{Var}\!\bigl[\bar{R}_\mathrm{owl}(S)\bigr]$.

Intermediate risk variance comparable to the
extreme chronotypes, implying that $\kappa=0$ is not sufficient to
insulate the agent from scenario-induced preference shifts.

\subsection{Analysis Results}
\label{sec:analysis}

All 27 conditions were simulated with $N_\mathrm{MC} = 20$ replicates
each. All six predictions are confirmed. Table~\ref{tab:pred_summary}
provides a compact summary; the subsections below discuss each in detail.

\begin{table}[H]
\centering
\caption{Prediction outcomes ($N_\mathrm{MC}=20$, centre position
         unless noted).}
\label{tab:pred_summary}
\medskip
\begin{tabular}{cllc}
\toprule
ID & Prediction & Key criterion & Outcome \\
\midrule
P1 & Epistemic invariance
   & Ambiguity gap $<0.10$~nat across chronotypes at S2
   & \checkmark \\
P2 & Risk sign-reversal
   & $\bar{R}_\mathrm{lark}(\mathrm{S3})>\bar{R}_\mathrm{lark}(\mathrm{S1})$
     and $\bar{R}_\mathrm{owl}(\mathrm{S1})>\bar{R}_\mathrm{owl}(\mathrm{S3})$
   & \checkmark \\
P3 & Spatial ordering at S2
   & centre $>$ near\_source $>$ far\_corner (legibility gradient)
   & \checkmark \\
P4 & Spatial ordering inverts at S3
   & far\_corner $>$ near\_source $>$ centre (complete reversal)
   & \checkmark \\
P5 & Position $\times$ chronotype at S3
   & Owl gap far\_corner$-$centre $\gg$ lark gap
   & \checkmark \\
P6 & Intermediate robustness
   & Lowest EFE risk variance across scenarios
   & \checkmark \\
\bottomrule
\end{tabular}
\end{table}

\subsubsection{P1 --- Epistemic Invariance Across Chronotypes}

At S2/centre ($\phi_\mathrm{text}=1.00$, $\phi_\mathrm{eye}=0.40$),
the performance ambiguity gap across chronotypes is $0.0014$~nat ---
more than two orders of magnitude below the $0.10$~nat threshold.
P1 is confirmed at every position and scenario tested. The near-zero
gap confirms that the epistemic channel is operationally chronotype-blind.

\subsubsection{P2 --- Risk Sign-Reversal Between Lark and Owl}

\begin{table}[H]
\centering
\caption{Mean session risk (nat) at centre ($N_\mathrm{MC}=20$).
         Bold = highest-risk scenario per chronotype.}
\label{tab:risk}
\medskip
\begin{tabular}{lccc}
\toprule
Chronotype & S1 & S2 & S3 \\
\midrule
Lark & 0.885 & 1.023 & \textbf{1.523} \\
Owl  & \textbf{1.352} & 0.679 & 0.623 \\
\bottomrule
\end{tabular}
\end{table}

The lark's risk nearly doubles from S1 to S3; the owl's approximately
halves. P2 is confirmed. The mechanism is the tolerance factor
$\kappa_c$: at S3 ($m>0$, over-arousal), the lark's $C_\mathrm{low}$
decreases (heightened aversion) while the owl's increases (reduced
aversion). The directions reverse at S1 ($m<0$).

\subsubsection{P3 --- Spatial Ordering Under S2}

\begin{figure}[H]
    \centering
    \includegraphics[width=0.9\linewidth]{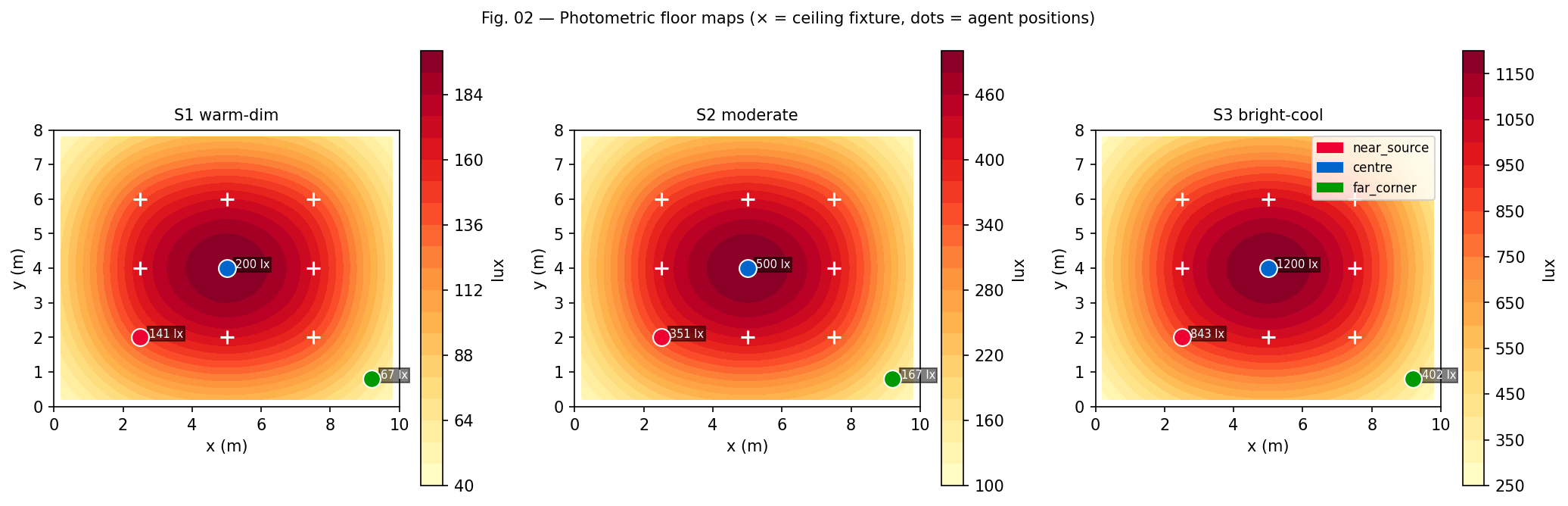}
    \caption{Photometric floor maps for the three lighting scenarios.
         Colour encodes illuminance in lux; white crosses mark ceiling
         fixture positions; coloured dots mark the three agent desk
         positions with their effective illuminance values.
         The spatial structure determines both precision functions:
         at S3, the far-corner position receives 402~lux
         ($\phi_\mathrm{text}=0.941$, $\phi_\mathrm{eye}=0.400$),
         while the centre position receives 1200~lux
         ($\phi_\mathrm{text}=0.047$, $\phi_\mathrm{eye}=0.002$).
         This twenty-fold difference in legibility precision, combined
         with the affective channel's over-arousal drift at S3, drives
         the complete inversion of the spatial performance ordering
         examined in P4: the far corner --- furthest from the fixtures
         --- outperforms the centre under intense cool light, despite
         being the worst-performing position under moderate light
         (see P3 and Table~\ref{tab:p3_wpm}).}
    \label{fig:placeholder}
\end{figure}

\begin{table}[H]
\centering
\caption{Session wpm under S2 by chronotype and position
         ($N_\mathrm{MC}=20$).}
\label{tab:p3_wpm}
\medskip
\begin{tabular}{lccc}
\toprule
Chronotype & \texttt{centre} & \texttt{near\_source} & \texttt{far\_corner} \\
\midrule
Lark         & \textbf{77.3} & 71.4 & 59.9 \\
Intermediate & \textbf{79.2} & 73.0 & 61.0 \\
Owl          & \textbf{82.8} & 75.1 & 64.8 \\
\bottomrule
\end{tabular}
\end{table}

\begin{figure}[H]
    \centering
    \includegraphics[width=0.9\linewidth]{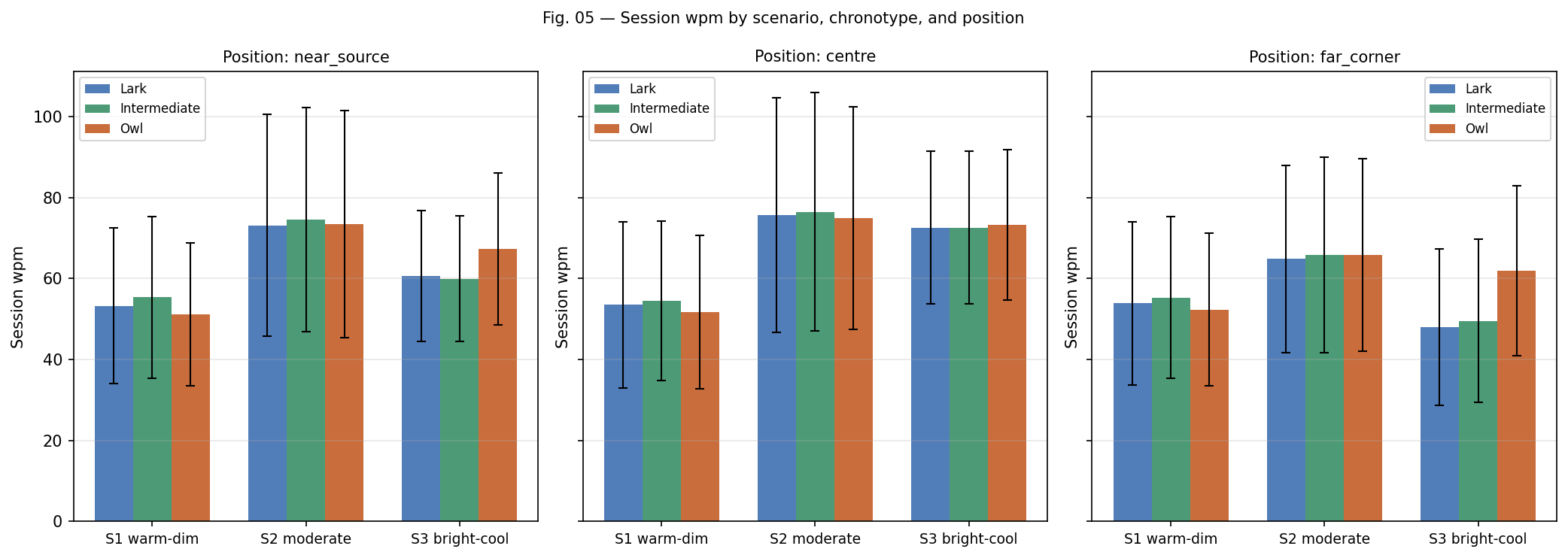}
    \caption{Session wpm by lighting scenario, chronotype, and spatial
         position ($N_\mathrm{MC}=20$; error bars are $\pm 1$~SD
         across replicates). Under S2 (centre panel), the spatial
         ordering
         \texttt{centre}~$>$~\texttt{near\_source}~$>$~\texttt{far\_corner}
         holds consistently for all three chronotypes (P3): the
         legibility gradient maps directly onto performance. Under S3
         (right panel), this ordering is completely reversed:
         \texttt{far\_corner}~$>$~\texttt{near\_source}~$>$~\texttt{centre}
         (P4). The position that performed worst under moderate lighting
         performs best under intense cool lighting. Under S1 (left
         panel), scenario-level differences are modest and spatial
         variation is small, reflecting the low and nearly uniform
         illuminance across all positions. P3 and P4 together constitute
         the critical experiment for the three-channel hypothesis
         (see Section~\ref{sec:predictions}).}
    \label{fig:placeholder}
\end{figure}

The spatial ordering centre $>$ near\_source $>$ far\_corner holds for
all three chronotypes. P3 is confirmed. Under S2, with zero Yerkes--Dodson
drift, the epistemic channel alone determines spatial variation: better
legibility produces better self-monitoring and more efficient action
selection.

\begin{figure}[H]
\centering
\includegraphics[width=0.72\textwidth]{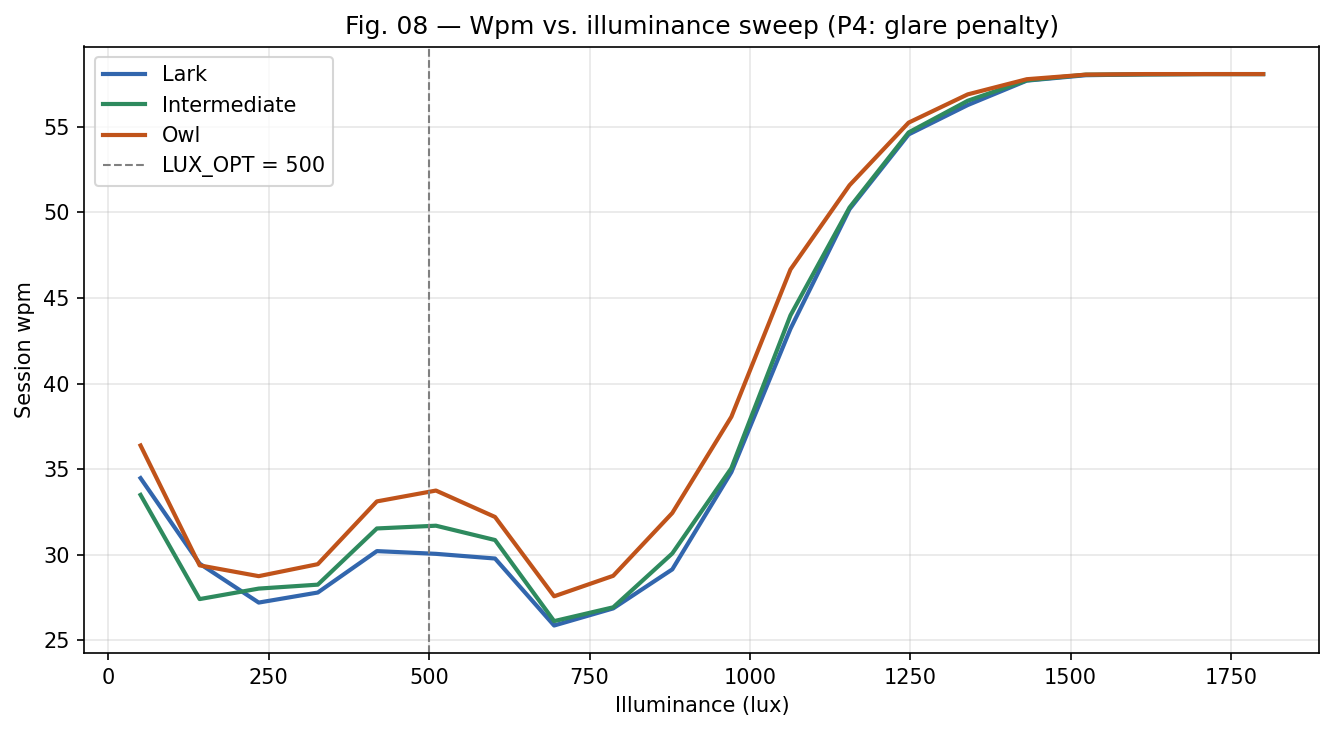}
\caption{Continuous lux sweep: session wpm as a function of illuminance
         (S2 scenario, centre position, $n=5$ seeds per lux level).
         The inverted-U shape reflects the text legibility precision
         function $\phi_\mathrm{text}(\ell)$: wpm rises steeply from
         50~lux to the optimum at $\ell^*=500$~lux (dashed line), then
         declines more gradually as glare accumulates. The asymmetric
         slope --- steeper on the under-illumination side than on the
         glare side --- reflects the mechanical legibility factor
         ($\ell_\mathrm{factor}=0.65+0.35\,\phi_\mathrm{text}$), which
         sets a floor on reading speed even at extreme illuminance.
         Chronotype differences are visible primarily around the
         optimum, where precision differences in $\gamma(t,c)$ amplify
         the EFE-minimising advantage of a well-matched arousal state.}
\label{fig:lux_sweep}
\end{figure}

\subsubsection{P4 --- Spatial Ordering Inverts Under S3}

\begin{table}[H]
\centering
\caption{Session wpm and EFE decomposition under S3 (MC mean across
         chronotypes, $N_\mathrm{MC}=20$). The spatial ordering is
         completely inverted relative to S2
         (Table~\ref{tab:p3_wpm}).
         $\ell_\mathrm{factor} = 0.65 + 0.35\,\phi_\mathrm{text}$.}
\label{tab:p4_s3}
\medskip
\begin{tabular}{lcccccc}
\toprule
Position & $\phi_\mathrm{text}$ & $\phi_\mathrm{eye}$
         & $\ell_\mathrm{factor}$ & wpm & Pause\%
         & Amb / Risk \\
\midrule
\texttt{far\_corner}
  & 0.941 & 0.400 & 0.979 & \textbf{71.3} & 32\%
  & 1.805 / 1.263 \\
\texttt{near\_source}
  & 0.479 & 0.108 & 0.818 & 63.7 & 29\%
  & 2.106 / 1.154 \\
\texttt{centre}
  & 0.047 & 0.002 & 0.666 & 59.9 & 18\%
  & 2.197 / 1.063 \\
\bottomrule
\end{tabular}
\end{table}

P4 is confirmed for all three chronotypes
(lark: far=67.6, near=60.0, centre=59.5;
 intermediate: far=69.8, near=62.6, centre=59.9;
 owl: far=76.4, near=68.5, centre=60.3).
The spatial ordering under S3 is the exact reverse of the ordering
under S2: \texttt{far\_corner} is best, \texttt{centre} is worst.

The EFE decomposition reveals the two-part mechanism. At
\texttt{far\_corner}/S3: both precision values are substantial
($\phi_\mathrm{text}=0.941$, $\phi_\mathrm{eye}=0.400$); total
ambiguity is low (1.805~nat); risk is high (1.263~nat), because
accurate self-monitoring reveals a predominantly deteriorated cognitive
state; the agent pauses $32\%$ of steps. But when it does work, the
mechanical legibility factor is $\ell_\mathrm{factor}=0.979$ ---
virtually no glare tax. At \texttt{centre}/S3: both precision values
are near zero ($\phi_\mathrm{text}=0.047$, $\phi_\mathrm{eye}=0.002$);
total ambiguity is high (2.197~nat); risk is moderate (1.063~nat),
because the near-flat $\mathbf{A}$ matrices make all predicted
observations nearly uniform, insensitive to the true (bad) state; the
agent pauses only $18\%$ of steps. But every working step is penalised:
$\ell_\mathrm{factor}=0.666$, reducing all output by one third.

The inversion arises from two forces operating in the same direction
at \texttt{far\_corner}: strategic rest (avoiding wasted effort in a
bad state) combined with mechanical efficiency (each working step is
nearly penalty-free). At \texttt{centre}, the opposite forces combine:
blind persistence (ignoring the bad state) combined with mechanical
penalty (taxing every working step). The position that supports strategic rest without imposing a severe mechanical glare penalty produces more output.

This result requires the multi-channel architecture for two distinct
reasons. First, the ambiguity-risk dissociation across positions ---
low ambiguity and high risk at \texttt{far\_corner}, high ambiguity and
moderate risk at \texttt{centre} --- requires the epistemic channel to
modulate self-monitoring quality by position. Second, the mechanical
legibility penalty requires the epistemic channel to act directly on
output speed per working step. Neither effect is present in an
arousal-only model, where desk position is irrelevant. Together, P3
and P4 constitute the critical experiment: the same epistemic channel
that produces a monotonic gradient under S2 (affective channel neutral)
produces a complete reversal under S3 (affective channel active).

\subsubsection{P5 --- Position-by-Chronotype Asymmetry Under S3}

\begin{table}[H]
\centering
\caption{Wpm advantage of \texttt{far\_corner} over \texttt{centre}
         under S3, by chronotype ($N_\mathrm{MC}=20$). Positive gap
         means far\_corner outperforms centre.}
\label{tab:p5_gap}
\medskip
\begin{tabular}{lrrrrc}
\toprule
Chronotype & \texttt{far\_corner} & \texttt{centre} & Gap
           & $\gamma$ & Pause\% at far \\
\midrule
Owl          & 76.4 & 60.3 & \textbf{+16.2} & 4.85 & 25\% \\
Intermediate & 69.8 & 59.9 & +9.9           & 4.93 & 33\% \\
Lark         & 67.6 & 59.5 & +8.1           & 3.88 & 36\% \\
\bottomrule
\end{tabular}
\end{table}

P5 is confirmed. The owl's far\_corner advantage ($+16.2$~wpm)
is twice the lark's ($+8.1$~wpm). The mechanism operates through two
concurrent effects of the tolerance factor $\kappa_c$ under over-arousal
($m>0$ at S3).

At \texttt{far\_corner}/S3, both observation channels are informative
($\phi_\mathrm{text}=0.941$, $\phi_\mathrm{eye}=0.400$) and the agent
accurately perceives its deteriorating state. For the lark
($\kappa=-3.0$), this accurate perception is costly: sharper aversion
to \texttt{low\_perf} inflates risk to $1.771$~nat and triggers
aggressive pausing ($36\%$ of steps). For the owl ($\kappa=+3.5$),
the same accurate perception is tolerated: reduced aversion keeps risk
at $0.772$~nat and pause rate at only $25\%$. The owl's higher policy
precision ($\gamma=4.85$ vs $3.88$) amplifies this advantage by making
it more efficient at exploiting the position's good epistemic conditions.

At \texttt{centre}/S3, both channels are suppressed and the two
chronotypes produce similar wpm ($59.5$ and $60.3$ respectively):
when neither agent can see its own state, the tolerance asymmetry has
little consequence.

The gap therefore arises entirely from what happens at
\texttt{far\_corner}: the owl's arousal tolerance makes it better able
to use accurate self-knowledge productively, while the lark's
intolerance of over-arousal turns the same accurate information into
a trigger for excessive rest.

\begin{figure}[H]
    \centering
    \includegraphics[width=0.9\linewidth]{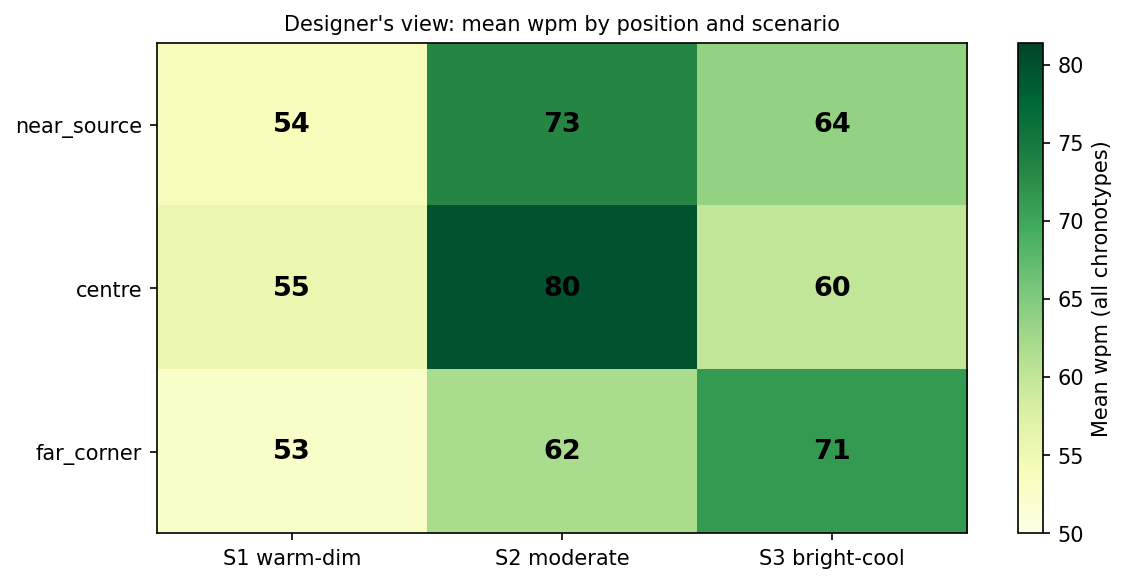}
    \caption{Designer's view: mean session wpm averaged across all three
         chronotypes, by spatial position (rows) and lighting scenario
         (columns). Numbers in each cell are wpm. The position-by-scenario
         interaction is clearly visible: under S2, the centre position
         dominates; under S3, far-corner and near-source positions
         recover relative to centre, with far-corner producing the
         highest population-level mean ($\approx 71$~wpm). The
         chronotype-specific asymmetry of this effect --- the subject of
         P5 --- is not visible in this population-average view but is
         reported in Table~\ref{tab:p5_gap}.}
    \label{fig:placeholder}
\end{figure}

\subsubsection{P6 --- Intermediate Chronotype Has Lowest Risk Variance}

\begin{figure}[H]
    \centering
    \includegraphics[width=0.9\linewidth]{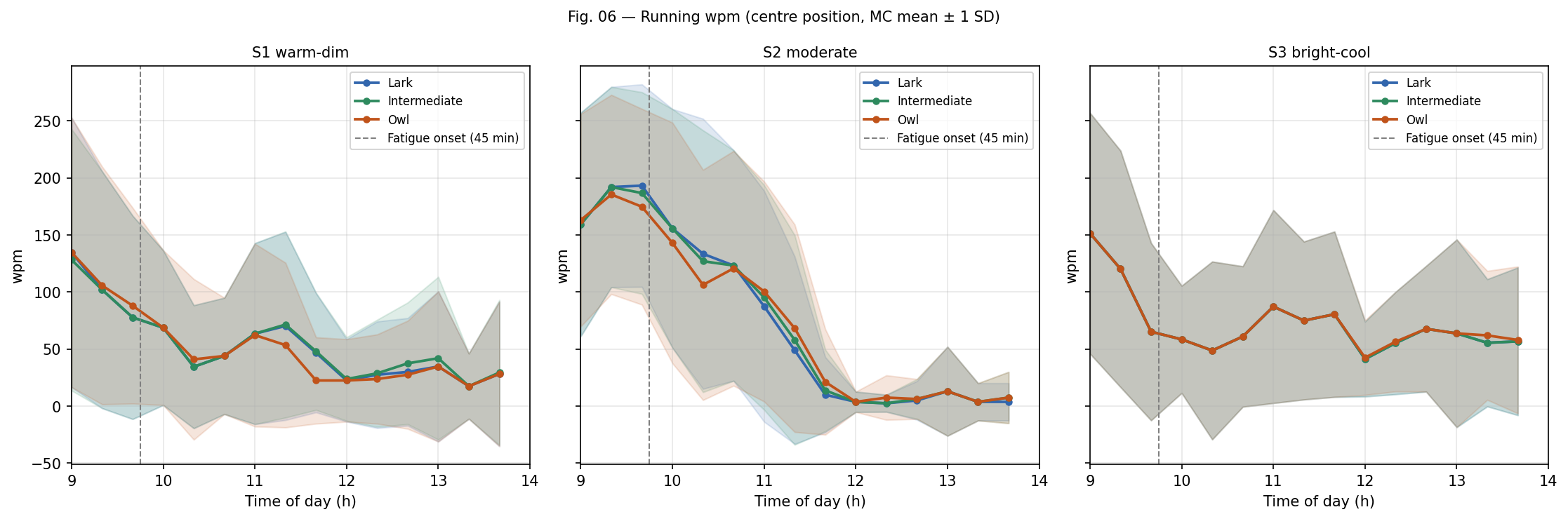}
    \caption{Running wpm over the $T=15$ decision steps (MC mean $\pm 1$~SD,
         centre position, $N_\mathrm{MC}=20$). Each panel shows one
         lighting scenario; coloured lines correspond to chronotypes.
         The dashed vertical line marks the fatigue onset at 45~minutes
         (step~3). In both S1 (left panel) and S3 (right panel), the
         lark--owl ordering is absent or reversed in the first half of
         the session and settles into its final direction only in the
         second half, consistent with P6. The large step-to-step
         variance reflects the stochastic nature of the true-state
         transitions and the coarse 20-minute temporal resolution.}
    \label{fig:placeholder}
\end{figure}

\begin{table}[H]
\centering
\caption{Variance of mean session risk across scenarios S1--S3
         (centre position, $N_\mathrm{MC}=20$).}
\label{tab:p6_riskvar}
\medskip
\begin{tabular}{lc}
\toprule
Chronotype & Risk variance \\
\midrule
Lark         & 0.0750 \\
Intermediate & \textbf{0.0111} \\
Owl          & 0.1098 \\
\bottomrule
\end{tabular}
\end{table}

\begin{figure}[H]
    \centering
    \includegraphics[width=0.9\linewidth]{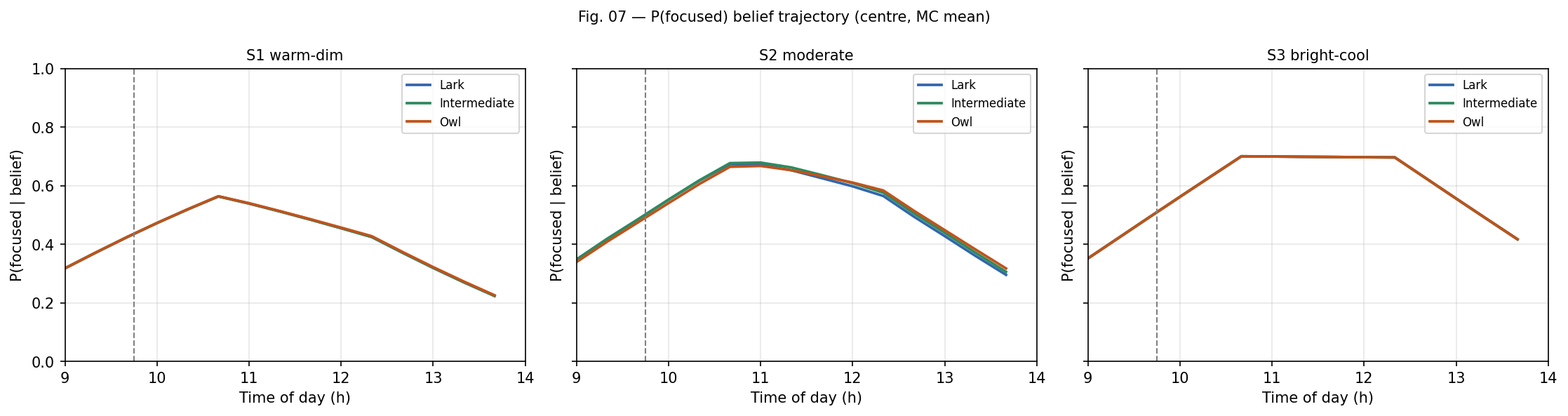}
    \caption{Posterior belief $Q(s=\texttt{focused})$ over the session
         (MC mean, centre position, $N_\mathrm{MC}=20$). The dashed
         vertical line marks the fatigue onset at 45~minutes (step~3).
         Three phases are distinguishable across all conditions:
         a stable high-belief phase (steps 0--2, $f(t)=1.00$);
         a progressive decline phase (steps 3--8, $f(t)$ falling
         from 0.93 to 0.46); and a near-collapsed phase (steps 9--14,
         $f(t)\leq0.34$). Under S3/centre, where both observation
         channels are suppressed ($\phi_\mathrm{text}=0.047$,
         $\phi_\mathrm{eye}=0.002$), the posterior remains close to
         the prior $\mathbf{D}$ throughout, compressing confidence bands
         and masking the genuine deterioration of the true state.}
    \label{fig:placeholder}
\end{figure}

P6 is confirmed. The intermediate's risk variance ($0.011$) is $6.8\times$
lower than the lark's and $9.9\times$ lower than the owl's. This is
a structural invariant of the model: with $\kappa_\mathrm{int.}=0$,
the mismatch $m(t,c)$ does not shift $C_\mathrm{low}$, making risk
independent of scenario by construction. This prediction would survive
any reparametrisation of the model as long as $\kappa_\mathrm{int.}=0$
is maintained.

\subsubsection{Fatigue as a Session-Level Moderator}

An unexpected finding concerns the relative wpm decline from the first
to the second half of the session:
\begin{equation*}
  \Delta\overline{\mathrm{wpm}}_{\mathrm{S1}} \approx 34,
  \qquad
  \Delta\overline{\mathrm{wpm}}_{\mathrm{S2}} \approx 100,
  \qquad
  \Delta\overline{\mathrm{wpm}}_{\mathrm{S3}} \approx 15.
\end{equation*}
The decline is largest under S2 and smallest under S3. The mechanism
is the same as P4: under S2 ($\phi_\mathrm{text}=1.00$,
$\phi_\mathrm{eye}=0.40$), the agent accurately perceives its
progressive fatigue and responds with appropriate but output-reducing
rest. Under S3/centre ($\phi_\mathrm{text}=0.047$,
$\phi_\mathrm{eye}=0.002$), both signals are blind to fatigue; the
agent keeps working but at reduced speed due to $\ell_\mathrm{factor}$.
The two-modality design makes the fatigue buffering effect of glare
visible at a finer level: not only does the agent fail to perceive
fatigue, it fails across \emph{both} observation channels simultaneously,
leaving it with no route to accurate self-assessment.

\section{Summary of Results}
\label{sec:summary}

The simulation supports the three-channel hypothesis and yields a coherent set of results. The epistemic channel is invariant across chronotypes: at any fixed desk position and lighting scenario, morning, intermediate, and evening types receive the same quality of information about their own state. By contrast, the affective channel produces a sign-reversal between morning and evening types: bright-cool light increases risk for morning types and decreases it for evening types, while warm-dim light produces the opposite pattern. Under moderate light (S2), performance follows the expected legibility gradient, with centre outperforming near-source and far-corner. Under intense cool light (S3), however, this ordering reverses: far-corner outperforms centre because it preserves informative observation signals and imposes little glare penalty, whereas centre suppresses both channels and mechanically slows reading. This advantage is larger for evening types than for morning types, while the intermediate chronotype remains the most stable across scenarios because its preference structure is less sensitive to arousal mismatch. An additional unexpected result concerns fatigue over time: the decline in reading speed is greatest under S2 and smallest under S3. The reason is that, under S2, the agent can accurately detect fatigue and responds by resting more, whereas under S3 glare suppresses both observation channels, so the agent continues working despite deteriorating performance. In this sense, glare acts as an involuntary fatigue buffer: not by improving functioning, but by reducing access to the evidence of decline.

\begin{figure}[H]
    \centering
    \includegraphics[width=0.9\linewidth]{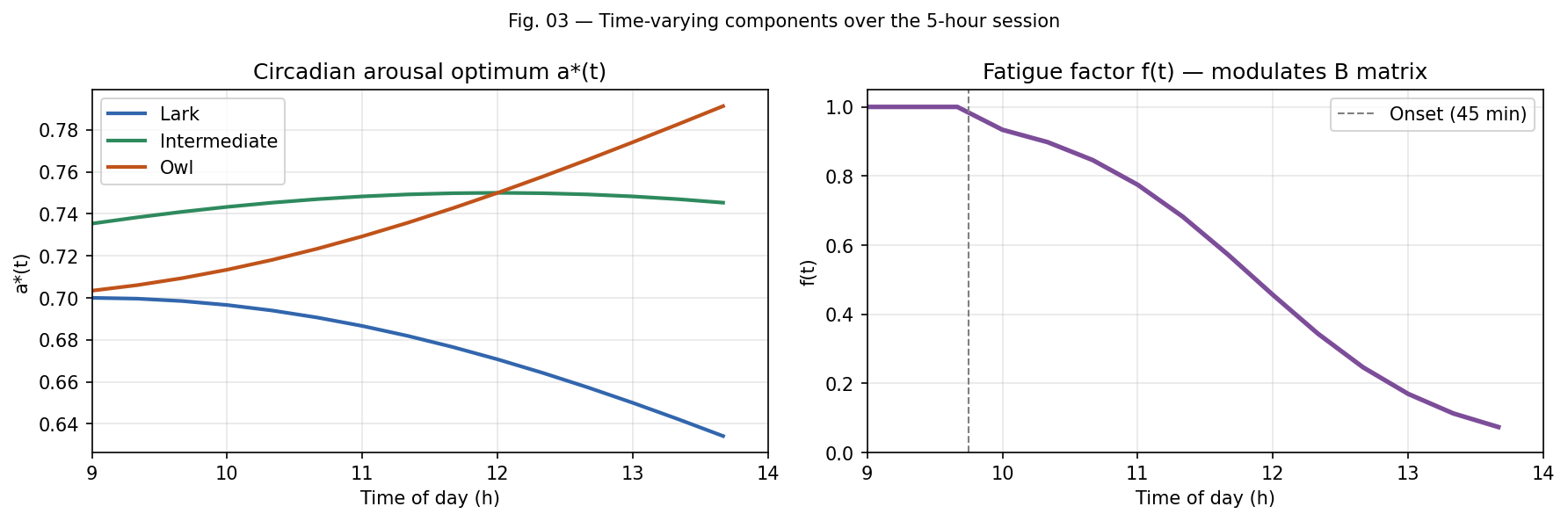}
    \caption{\emph{Left}: Circadian arousal optima $a^*(t,c)$ for the three
         chronotypes over the five-hour session (09:00--14:00). The lark
         (blue) declines from $0.70$ at session onset to $0.63$ by
         14:00; the owl (orange) rises from $0.70$ to $0.80$; the
         intermediate (green) follows a shallow arc peaking near noon.
         The $0.17$-unit divergence between lark and owl by session end
         drives the half-session crossover.
         \emph{Right}: Fatigue factor $f(t)$. Full attentional capacity
         ($f=1.00$) is maintained for the first 45~minutes; thereafter
         $f$ follows a sigmoid decay, reaching $f=0.46$ at 160~minutes
         and $f=0.07$ at the end of the session. The fatigue factor
         modulates $\mathbf{B}(t,a)$ via linear interpolation between
         the fresh and tired transition matrices.}
    \label{fig:placeholder}
\end{figure}

 \section{Objections and Replies}
\label{sec:objections}

\subsection*{O0. The predictions are post-hoc rationalisations of simulation outputs.}

An objector may argue that the six predictions were formulated only after the model had already been built and simulated, and therefore amount to redescriptions of the output rather than genuine predictions.

\paragraph{Reply.}
This objection misstates the logic of the paper. The predictions are not summaries of whatever the simulation happened to produce. They follow from specific structural commitments made when the generative model was defined. The derivational order is: hypothesis, architecture, prediction, simulation. The role of the simulation is to verify that those architectural commitments generate the expected qualitative signatures.

This distinction matters because each prediction probes a different part of the model. P1 tests whether the epistemic channel remains independent of chronotype. P2 tests whether the affective channel generates the expected reversal between morning and evening types. P3 and P4 test whether epistemic and affective mechanisms interact in a way that produces both a monotonic spatial gradient under S2 and its reversal under S3. P5 tests whether chronotype and spatial position interact through the preference structure. P6 tests whether setting $\kappa_{\mathrm{int.}}=0$ is sufficient to make the intermediate chronotype comparatively invariant across scenarios. In each case, a failure would implicate a specific structural commitment rather than a merely cosmetic choice of wording.

\subsection*{O1. The parameters are calibrated to make the predictions come out right.}

A second objection is that several parameter values are not empirically estimated and may have been chosen so that the six predictions hold. On this view, the model shows only that one internally coherent parametrisation exists.

\paragraph{Reply.}
This objection is substantially correct, and the paper does not deny it. The present model is not offered as an empirically calibrated account of human reading performance. It is a proof of concept designed to show that the three-channel hypothesis can be implemented as a coherent generative architecture that yields distinct, falsifiable predictions. At this stage, the main question is not whether the chosen values are already correct in an empirical sense, but whether the architecture is tractable and whether its qualitative behaviour is robust enough for later calibration to be meaningful.

For that reason, the strongest support at this stage comes from structural robustness rather than empirical fit. Appendix B shows that the model's main qualitative signatures remain interpretable across plausible variations of several key parameters. In addition, some modelling choices are not introduced as fitted values at all, but as theoretical commitments. The intermediate chronotype is assigned $\kappa_{\mathrm{int.}}=0$ because the model treats it as having no systematic arousal-tolerance asymmetry. The text-legibility precision function is symmetric because there is no clear reason to assume that dim light and glare should affect text legibility asymmetrically in the same way as the oculomotor signal. By contrast, the eye-tracking precision function is asymmetric because glare has a specific physiological effect through pupil constriction that is not mirrored under low illuminance. These choices are therefore open to empirical challenge, but they are not merely tuned to rescue the model.

\subsection*{O2. The three channels are not structurally independent.}

A third objection is that the channels are not genuinely distinct because they share upstream physical drivers. In particular, the spectral coefficient $\mu(\sigma)$ contributes both to the affective channel, through baseline arousal, and to the normative channel, through the policy prior.

\paragraph{Reply.}
This is a real structural feature of the model, but it is not a defect. The channels are distinguished by the component they modify and by the computational role that component plays in inference and action selection, not by complete independence of physical input. A single environmental variable can affect what the agent can observe, what the agent prefers, and what the agent is disposed to do without collapsing those functions into a single mechanism.

For this reason, the relevant test of channel distinctness is not input purity but selective ablation. If one removes the spectral contribution to arousal while leaving the normative prior intact, the affective signature should weaken while the normative effect should remain. That is the relevant sense in which the channels are structurally distinct: they can be functionally dissociated even when they are not driven by wholly separate physical variables.

\subsection*{O3. A simpler, single-channel model would explain the same results.}

An objector may further claim that the three-channel architecture is unnecessary. A suitably chosen one-factor model, for example an arousal-only model, might reproduce the main scenario-by-chronotype pattern more parsimoniously.

\paragraph{Reply.}
The strongest answer to this objection is the joint force of P3 and P4. Under S2, when the affective channel is at its neutral point, performance follows the legibility gradient: centre performs best, followed by near-source, then far-corner. Under S3, this ordering reverses: far-corner outperforms centre. A model in which lighting affects only arousal has no reason to predict a position effect within a fixed scenario, since arousal is determined by the scenario rather than by desk location. A model in which lighting affects only legibility predicts the same spatial ordering in both scenarios, because better legibility should always improve output. Neither kind of single-channel model predicts a pattern that is monotonic under S2 and reversed under S3.

The same point appears in the EFE decomposition. Under S3, far-corner combines lower ambiguity with higher risk, whereas centre combines higher ambiguity with lower risk. That dissociation matters because it shows that the better-performing position is not the one with the easier decision problem. A single-channel model has no straightforward way to represent this combination of improved self-monitoring, increased aversiveness of the detected state, and reduced mechanical glare cost within the same scenario. The three-channel architecture is therefore not merely sufficient; it is doing explanatory work that the simpler alternatives do not obviously reproduce.

\subsection*{O4. The agent is cognitively unrealistic.}

Another objection is that the agent is too idealised. It plans ahead, updates beliefs with exact Bayesian inference, has no episodic memory or external distraction, and experiences fatigue only through a predetermined function.

\paragraph{Reply.}
This objection is valid only if the model is read as a descriptive simulation of an actual human reader. That is not the claim of the paper. The model is normative: it specifies how an agent with a particular generative model and preference structure would behave under changing lighting conditions. In active inference, this idealisation is useful rather than embarrassing, because it allows empirical deviations to become informative about what the model leaves out.

The simplifications are also not arbitrary. The planning horizon is meant to capture the timescale of decisions about continuing, rereading, or pausing. Policy precision already introduces a graded form of bounded rationality, since low precision makes action selection less tightly coupled to expected free energy. The addition of the oculomotor channel also weakens an important idealisation present in simpler models: it does not assume that the agent always has access to a reliable self-performance signal. Under glare, even this second source of information degrades, introducing bounded observability into the model itself.

\subsection*{O5. The model adds nothing to the Yerkes--Dodson curve.}

A final objection is that the model simply redescribes a familiar inverted-U relation between arousal and performance in more formal language.

\paragraph{Reply.}
The Yerkes--Dodson curve is a descriptive regularity, not a computational architecture. It says that performance is often best at intermediate arousal, but it does not specify which part of the decision process is being affected, how self-monitoring enters the picture, how spatial differences in illuminance interact with arousal, or how those effects unfold over time for different chronotypes. The present model addresses exactly those questions.

In that sense, the paper does not merely restate the inverted-U relation. It embeds it within a richer structure that distinguishes perceptual precision, affective preference, and normative action bias, and then derives predictions that a bare arousal-performance curve cannot express. The clearest example is P4: under specific spatial and affective conditions, better self-monitoring leads to more rest and higher total output, whereas severe glare leads to more persistence but lower output. A descriptive inverted-U curve has no observer model and no mechanism for that result. If predictions of this kind are confirmed empirically, then the familiar curve becomes a limiting case of a more informative computational account rather than a sufficient explanation in its own right.

\section{Conclusions}
\label{sec:conclusions}

This paper develops a proof-of-concept active inference model of the cognitive and emotional effects of shared indoor lighting. Its main contribution is not to provide a calibrated account of human reading behaviour, but to formalise the hypothesis that lighting shapes behaviour through three computationally distinct but interacting channels: an epistemic channel affecting self-monitoring precision, an affective channel affecting the evaluation of performance under different arousal states, and a normative channel affecting action tendencies. In this sense, the value of the model lies less in any single numerical result than in showing that the three-channel hypothesis can be expressed as an explicit generative architecture with distinct, falsifiable behavioural signatures.

The most important next step is therefore empirical. The model specifies what should be measured --- reading speed, oculomotor metrics, pause frequency, and subjective confidence --- and under what conditions these measures become theoretically informative: prolonged reading sessions, multiple desk positions, different lighting scenarios, and mixed chronotype groups. On this view, the strongest test of the framework is not whether a single aggregate performance curve is replicated, but whether the predicted dissociations emerge across measures: for example, whether chronotypes differ in risk across scenarios, whether intermediate types are more stable than extreme chronotypes, and whether behavioural output diverges from self-monitoring under severe glare. If such signatures are confirmed, indoor lighting will need to be understood not only as a variable that changes performance, but also as a condition that shapes how agents perceive themselves, evaluate their own state, and regulate action over time.

\appendix

\section{Appendix. Literature Review Analysis}
\label{app:litreview}

This appendix focuses on the results of the literature review. Table 5 provides a detailed overview of the studies included in the literature review, summarizing key characteristics such as study design, sample, experimental conditions, and main findings. It is intended to complement the conceptual synthesis presented in the main text and to ensure transparency and completeness of the review. The papers are presented in chronological order.

\begin{landscape}
 
\begin{footnotesize}
\setlength{\LTcapwidth}{\linewidth}
 
\begin{longtable}{P{2cm} P{6.2cm} P{4.8cm} P{4.8cm} P{5.2cm}}
 
\caption{Summary of reviewed studies on the cognitive and emotional effects of illuminance (lux), correlated colour temperature (Kelvin), and their combined interactions.}
\label{tab:litreview} \\
 
\toprule
\textbf{Paper} &
\textbf{Study design} &
\textbf{Cognitive effects of illuminance (lux)} &
\textbf{Cognitive effects of CCT (Correlated Colour Temperature, Kelvin)} &
\textbf{Combined effects of K-lux} \\
\midrule
\endfirsthead
 
\multicolumn{5}{l}{\small\textit{(Continued from previous page)}} \\[4pt]
\toprule
\textbf{Paper} &
\textbf{Study design} &
\textbf{Cognitive effects of illuminance (lux)} &
\textbf{Cognitive effects of CCT (Correlated Colour Temperature, Kelvin)} &
\textbf{Combined effects of K-lux} \\
\midrule
\endhead
 
\midrule
\multicolumn{5}{r}{\small\textit{(Continued on next page)}} \\
\endfoot
 
\bottomrule
\endlastfoot
 
Knez (2001) &
108 paid 18-year-old students (equal gender) participated in a laboratory study in a neutral, temperature-controlled room (21\,°C) under three lighting conditions (3000, 4000, 5500\,K). They completed mood recognition, room evaluation, problem-solving, and short- and long-term memory tasks following extended reading. &
--- &
The variation of Kelvin did not have significant effects on the emotional level. On a cognitive level, short-term recall was best under warm light (3000\,K). Long-term recall showed gender differences: women performed best under daylight (5500\,K), while men performed better under warm (3000\,K) and cool (4000\,K) light. Problem-solving performance was highest under warm light (3000\,K), suggesting benefits for tasks requiring selective attention. &
--- \\[6pt]
 
Naylor \& Firth (2008) &
Experimental, within-subject repeated-measures design with 20 normally sighted university students (mean age $\approx$21). Participants completed the Wilkins Rate of Reading Test (WRRT) under four conditions: natural light ($\approx$1200\,lux) vs.\ artificial fluorescent light ($\approx$450\,lux) $\times$ text on white vs.\ pink paper. &
Participants read significantly faster under artificial lux lighting than under natural lux lighting (mean increase 4.62\,wpm; $\approx$2.7\% faster). Reading rate was 177.1\,wpm at 450\,lux vs.\ 172.5\,wpm at 1200\,lux, suggesting that very high illuminance may be slightly detrimental to reading speed, possibly due to glare or excessive page luminance. &
CCT was not explicitly manipulated or reported; the contrast is primarily between daylight (1200\,lux) and fluorescent artificial light (450\,lux). &
--- \\[6pt]
 
Chellappa et al.\ (2011) &
16 healthy young men ($\approx$24 years). Within-subject, balanced cross-over laboratory study (3 sessions, 1 week apart). Evening/night protocol. Lighting conditions: 6500\,K (blue-enriched CFL), 2500\,K (warm CFL), 3000\,K (incandescent). 2-hour light exposure after dark adaptation. Measures: salivary melatonin, sleepiness, well-being, visual comfort, mental effort; PVT, GO/NOGO, PVSAT. &
Illuminance was kept constant ($\sim$40\,lux) across all conditions. The study shows that even very low light levels can strongly affect alertness, melatonin, and cognitive performance if spectral composition is blue-enriched. &
6500\,K produced much stronger melatonin suppression, less sleepiness, higher well-being, better visual comfort, and faster reaction times compared to 2500\,K and 3000\,K. No significant effects on executive function. &
Because lux was fixed, the paper does not test a lux $\times$ Kelvin interaction. However, it demonstrates that spectral composition can dominate over intensity: low lux + high CCT (blue-enriched) can be cognitively and biologically very powerful. \\[6pt]
 
Barkmann et al.\ (2012) &
Quasi-experimental field study in two schools over nine months. Variable lighting (Philips SchoolVision) vs.\ standard lighting controls. Lighting scenarios: Standard (300\,lux, 4000\,K), Concentrate (1060\,lux, 5800\,K), plus Activate, Relax, etc. Standardised concentration test (d2), reading speed/comprehension, and questionnaires. &
--- &
--- &
Under the Concentrate program (1060\,lux, 5800\,K), students made 20.8\% fewer omission errors on the d2 test and read 16.8\% more words compared to Standard lighting. Reading comprehension showed a positive trend but did not reach significance. High illuminance combined with cooler CCT substantially boosted attention and reading speed. \\[6pt]
 
Mott et al.\ (2012) &
Quasi-experimental study with four third-grade classrooms across an entire school year. Conditions: Focus (1000\,lux, 6500\,K) vs.\ Normal (500\,lux, 3500\,K). Oral reading fluency (ORF) measured at beginning, mid-year, and end-year; motivation and concentration assessed via questionnaires and d2 test. &
--- &
--- &
A significant lighting $\times$ time interaction showed greater ORF gains under 1000\,lux/6500\,K, maintained across the year. No effects emerged for concentration or motivation, indicating benefits specific to reading fluency. \\[6pt]
 
Park et al.\ (2013a) &
32 healthy adults (16 female, mean age $\approx$26.8). Within-subject, repeated-measures. Task: Raven's Progressive Matrices. Four conditions: WL (2766\,K, 300\,lux), CL (5918\,K, 300\,lux), WH (2766\,K, 600\,lux), CH (5918\,K, 600\,lux). Frontal and occipital EEG alpha power; subjective valence and arousal ratings. &
Higher brightness caused significantly higher valence (people felt happier). Occipital alpha power was lower in high light than low light, suggesting brightness affects visual cortex activation. Brightness modulates emotional state and visual cortical activity, but does not show a simple ``more light = better cognition'' effect. &
Warm light (low\,K) made subjects feel more relaxed (lower arousal). Colour temperature did not significantly affect valence, only arousal. No strong main effect of CCT alone on alpha power. &
Highest alpha power (better focused internal processing) occurred in congruent combinations: warm + low light (WL) and cool + high light (CH). Lower alpha power in mismatched conditions (WH, CL). Cognitive brain state is determined by the combination of brightness and CCT, not by either alone. \\[6pt]
 
Park et al.\ (2013b) &
22 adults (11 female, mean age $\approx$23). Conditions: 3000\,K vs.\ 7100\,K; 150\,lux vs.\ 700\,lux (cool–dark, cool–bright, warm–dark, warm–bright). Sternberg working memory task (set sizes 3, 5, 7). EEG: frontal theta power; occipital ERPs (P1/N1). &
Higher illuminance (700\,lux) reduced frontal theta power during working memory retention (interpreted as a change in mental effort) and increased N1 amplitude. Brightness altered early attentional/sensory processing but did not translate into better or worse task performance. &
CCT had minimal/no significant effects across main outcomes. Key EEG effects were attributed primarily to illuminance. &
Lighting effects are largely explained by lux; no strong evidence of a meaningful lux $\times$ Kelvin interaction on behavioural performance. Mood/emotion changes were not measured. \\[6pt]
 
Keis et al.\ (2014) &
58 high-school students from two schools split into blue-enriched vs.\ normal lighting groups. Four conditions combining normal and blue-enriched light during lessons and tests (normal/normal, normal/blue, blue/normal, blue/blue). &
--- &
--- &
Under 5500\,K and 300\,lux lighting, cognitive performance increased. Full blue-enriched light exposure improved processing speed, concentration, and visual memory; students reported greater appreciation for this condition. \\[6pt]
 
Huiberts et al.\ (2015) &
64 students (32 male, 32 female; mean age 21.4). Two counterbalanced lab sessions in an office-like setting. Two illumination levels (200 and 1000\,lux) tested on digit span and n-back tasks. Participants completed four 15-min blocks after a 100-lux baseline. &
Higher light (1000\,lux) improved speed and accuracy on easy tasks; lower light (200\,lux) led to more stable performance on difficult tasks. More light does not necessarily improve performance; illumination should be matched to task demands. &
Held constant at 4000\,K — no effects. &
--- \\[6pt]
 
Choi \& Suk (2016) &
Series of lab + field experiments in elementary-school classrooms using tunable LEDs at three presets: 3500\,K (``Relax''), 5000\,K (``Standard''), 6500\,K (``Energy''), all at $\sim$500--600\,lux at desk level. Pre-test–post-test control-group design plus lab tests. &
Illuminance was intentionally kept in a narrow band ($\sim$450--570\,lux); the study effectively isolates CCT effects. &
6500\,K produced the greatest number of correct answers in arithmetic tests. 3500\,K was associated with higher comfort and recommended for rest/relaxation. 5000\,K suggested as default for reading and typical classroom activities. &
Within $\approx$500\,lux, CCT clearly differentiated functions: 3500\,K for rest; 5000\,K for ordinary reading/lecture; 6500\,K for intensive cognitive work with improved performance. Because illuminance was held roughly constant, the main conclusion is about CCT. \\[6pt]
 
Hartstein et al.\ (2018) &
38 healthy preschool children (4.5--5.5 years; 21 female). Windowless lab; tasks on a laptop. Tunable LED ceiling luminaires. Participants assigned to experimental (3500 to 5000\,K) or control (constant 3500\,K) groups. Go/No-Go and Hearts \& Flowers tasks assessed attention, inhibition, and cognitive flexibility. &
Illuminance was held approximately constant. &
Because lux was controlled, the study mainly isolates CCT effects. Go/No-Go: no lighting effect on accuracy; reaction time improved similarly in both groups (practice effect). Hearts \& Flowers (task switching): higher CCT (5000\,K) produced greater improvement in switch-trial accuracy than remaining at 3500\,K. &
With brightness held constant, changing CCT from 3500 to 5000\,K was enough to improve task-switching accuracy on switch trials. \\[6pt]
 
Ram \& Bhardwaj (2018) &
40 young adults ($\sim$20 years; 20 female). Four light sources at controlled 400\,lux: CFL ($\sim$3400\,K), fluorescent tube ($\sim$3000\,K), tungsten incandescent, and LED. Reading rate, contrast sensitivity, colour vision, and subjective comfort assessed after each condition. &
All conditions used 400\,lux; this paper does not test different lux levels — effects are purely spectral/source-type at constant illuminance. &
Because the CCT range is narrow ($\approx$3000--3400\,K) and multiple lamp characteristics change simultaneously (spectrum shape, flicker, CRI), effects on reading appear driven more by light-source quality/spectrum than by CCT as an abstract parameter. &
--- \\[6pt]
 
Zhu et al.\ (2019) &
Laboratory, 2 (illuminance: 200 vs.\ 1200\,lux) $\times$ 2 (CCT: 3000 vs.\ 6500\,K) $\times$ 2 (time of day: morning vs.\ afternoon) mixed design with 60 healthy young adults (30 female). Tasks: Go/No-Go, 2-back, facial expression recognition, long-term word recognition; mood (PANAS) and alertness ratings. &
Bright light (1200\,lux) made participants less sleepy and more positively toned than dim light (200\,lux). For Go/No-Go and 2-back, reaction times were slowest in the low-illuminance warm condition, and bright light improved performance. &
Go/No-Go accuracy showed a main effect of CCT: participants were less accurate in cool (6500\,K) than warm (3000\,K) light. For 2-back, CCT mainly entered through interactions with illuminance. &
The worst cognitive performance consistently appeared in low-illuminance + warm light (200\,lux, 3000\,K). Best long-term word recognition occurred in high-illuminance cool conditions (1200\,lux, 6500\,K) for neutral words in the afternoon. Bright light is reliably beneficial for alertness, mood, inhibition, and working memory; CCT modulates these effects. \\[6pt]
 
Grant et al.\ (2021) &
Randomized, within-subjects trial with sleep-restricted college-aged adults exposed to four polychromatic white LED spectra differing in short-wavelength enrichment (melanopic EDI) at roughly similar photopic illuminance. Cognitive battery: working memory/processing speed, motor sequence learning, vigilance (PVT), declarative memory; subjective alertness. &
Lux was held constant. &
Higher melanopic EDI ($\approx$ higher CCT) produced lower subjective sleepiness and better performance on some cognitive measures than low-melanopic ($\approx$ lower CCT) conditions. &
--- \\[6pt]
 
Ru et al.\ (2021) &
Within-subjects lab study with healthy adults tested on two separate days under low ($\sim$100\,lux) vs.\ high ($\sim$1000\,lux), both at $\approx$6500\,K. Tasks: sustained attention (PVT), response inhibition (Go/No-go), conflict monitoring (Flanker), working memory (2-Back, PVSAT); repeated measures of sleepiness, mood, and light appraisals. &
High vs.\ low illuminance did not significantly affect PVT or Flanker. For Go/No-go, high illuminance sped responses only for the easy version. For working memory, 1000\,lux improved accuracy and speed on 2-Back and accuracy on PVSAT; benefits were larger for easier trials. &
CCT was held constant (6500\,K); this study cannot speak to Kelvin effects directly. &
In a bright-cool office (6500\,K), raising illuminance from 100 to $\sim$1000\,lux selectively benefits working-memory and simple response-inhibition performance, especially for less difficult task versions. No uniform ``bright is always better'' effect; intellectual/working-memory tasks can profit from higher lux. \\[6pt]
 
Shishegar \& Boubekri (2022) &
Field, crossover intervention with 21 healthy older adults (mean age $\approx$77). Two 9-day lighting interventions separated by a 2-week washout. Whole-day dynamic scheme: blue-enriched morning light decreasing to warm, dim evening light. Cognitive performance (TMT, DSST) and mood (GDS, PANAS) assessed. &
Increasing daytime illuminance to 200--500\,lux (from chronically dim) produced significant improvements in mood and cognitive performance (faster TMT, better DSST). Moving to appropriately bright daytime lighting benefits older adults' attention, processing speed, and global cognition. &
Adding spectrum tuning (high CCT morning $\to$ neutral afternoon $\to$ low CCT evening) produced significantly larger improvements in both mood and cognitive measures than illuminance changes alone. &
High lux + high CCT in the morning supports better daytime cognitive functioning and mood. Variable illuminance alone was beneficial, but variable illuminance + variable CCT was better, indicating that Kelvin modulates the cognitive benefits of lux across the day. \\[6pt]
 
Castilla et al.\ (2023) &
Immersive VR replica of a university classroom; geometry, materials, and CCT held constant. Only illuminance manipulated at 100, 300, and 500\,lux on the desk plane. 142 university students completed memory tasks while HRV, EEG, and EDA were recorded. &
Memory performance was highest at 100\,lux, with significant declines at 300 and 500\,lux (which did not differ from each other). Neurophysiological markers showed higher activation at 100\,lux, mirroring better memory; higher illuminance was associated with reduced arousal/engagement. &
CCT was held fixed. &
--- \\[6pt]
 
Chauca et al.\ (2023) &
Experimental + review-based study on classroom LED lighting combining technical characterisation and student performance questionnaires to identify optimal lux and CCT. Integrates and interprets results from several prior empirical case studies. &
$\approx$400\,lux associated with optimal long-term memory. Near-linear increase in attention with illuminance, reaching notably higher attention at $\sim$1060\,lux compared with 300\,lux. &
6500\,K is repeatedly associated with higher learning, especially for arithmetic problem solving. &
6500\,K combined with 500--600\,lux yields more correct answers in arithmetic; $\approx$400\,lux appears optimal for long-term memory; $\sim$1000--1060\,lux for peak attention. The paper argues for differentiated settings: moderate lux for memory, higher lux for attention, cool CCT for intensive learning. \\[6pt]
 
Nole Fajardo et al.\ (2023) &
Immersive VR study comparing four classroom configurations differing in lighting, wall colour, and geometry. Lighting manipulations involved different levels of illuminance and CCT within realistic classroom ranges. University students performed attention and memory tasks; preferences collected with analyses segmented by gender. &
--- &
--- &
Rather than a monotonic rule, the study shows that changing illuminance and CCT meaningfully modulates attention and memory, with gender-specific sensitivities. The work strongly supports prioritising lighting (lux + Kelvin) over other spatial features for cognitive optimisation. \\[6pt]
 
Zhou \& Pan (2023) &
Full-scale mock classroom, 3 illuminance levels (300, 500, 750\,lux) $\times$ 3 CCTs (2700, 4000, 6500\,K) $\times$ 3 reading durations (15, 30, 60\,min). Twelve young adults performed a paper-based number-correction task simulating reading while EEG was recorded. &
--- &
--- &
Higher illuminance and higher CCT improved reading efficiency and reduced EEG-defined fatigue, but the optimal combination depended on reading duration: (a) 15\,min: best at 500\,lux \& 6500\,K; (b) 30\,min: best at 500\,lux \& 4000\,K; (c) 60\,min: best at 750\,lux \& 6500\,K. Lux and Kelvin interact with task duration; there is no single static optimal setting. \\[6pt]
 
Mostafavi et al.\ (2024) &
VR-based office simulation with five lighting conditions: 2000\,K (Warm-Dark, Warm-Bright), 4600\,K (Neutral-Moderate), 7200\,K (Cold-Dark, Cold-Bright); illuminance $\approx$100, 400, 700\,lux at workplane. Tasks: working memory (BDST) and visual memory (VMT) during morning, afternoon, and evening sessions. &
Brighter conditions ($\sim$700\,lux) generally yielded better scores on some memory tasks (especially VMT) than dark ($\sim$100\,lux). Effects were task- and time-of-day-dependent; benefits of higher illuminance were clearer for VMT and at specific times. &
Very warm light (2000\,K) tended to be disadvantageous for demanding working-memory performance, particularly in the afternoon. Neutral 4600\,K frequently produced stable, relatively good performance with moderate brightness (400\,lux). &
No single optimal lux--Kelvin pair. Broad pattern: bright and neutral/cool conditions ($\approx$400--700\,lux, $\geq$4600\,K) supported working and visual memory better than dim, very warm light ($\approx$100\,lux, 2000\,K), especially later in the day. Illuminance has a somewhat stronger and more consistent influence than CCT; CCT modulates the effect. \\[6pt]
 
Jung et al.\ (2024) &
Laboratory study in an artificial climate chamber with 16 healthy adults in their 20s. Nine office-type LED conditions: 3 CCTs (4000, 5000, 6500\,K) $\times$ 3 illuminances (200, 500, 800\,lux). Cognitive battery (primary and complex functions), psychophysiological measures, and subjective workload and fatigue; mixed linear models and regression. &
Within the realistic office range (200--800\,lux), higher illuminance was associated with better primary cognitive performance, especially attention. &
Higher CCT (toward 6500\,K) significantly improved memory and some executive-function measures when illuminance was held constant. &
Lux and Kelvin reliably enhance lower-level cognitive processes (attention, basic memory) within normal office ranges, but do not guarantee improvements in higher-order cognition or psychophysiological indicators. \\[6pt]
 
Li \& Yao (2025) &
Within-subject 3 $\times$ 3 design: 3 illuminance levels (200, 500, 1000\,lux) $\times$ 3 CCTs (2730, 3930, 5870\,K) with tunable LED ceiling luminaires in a simulated office. Healthy young adults read texts of two difficulty levels on a monitor; eye tracker recorded gaze metrics; Score of Reading Efficiency (SRE) computed; brightness, visual comfort, and pupil size measured. &
SRE increased monotonically from 200 $\to$ 500 $\to$ 1000\,lux in both easy and difficult texts; significantly better reading efficiency at 1000 vs.\ 200\,lux and typically at 500 vs.\ 200\,lux. &
CCT alone did not significantly change SRE or comfort in main-effect analyses, although comfort votes tended to rise with higher CCT (more participants rating ``comfortable'' at 5870\,K than at 2730\,K). &
Lux and CCT interacted: the positive effect of higher illuminance on SRE became more pronounced at higher CCTs (1000\,lux at 5870\,K yielded the highest reading efficiency). Higher illuminance and higher CCT produced smaller pupil sizes, negatively correlated with SRE, suggesting that bright, relatively cool light supports more efficient visual–cognitive processing during reading. \\[6pt]
 
Mott et al.\ (2025) &
High-school field experiment in four classrooms (students randomly assigned by grade-level strata). Blinds closed; illuminance and CCT carefully measured at desk and eye level. Energy (intervention): $\approx$4120\,K, $\sim$436\,lux, with higher spectral concentration in the blue-to-cyan band (440--520\,nm). Control: $\approx$3789\,K, $\sim$486\,lux, normal blue content. Students completed a practice ACT Reading Comprehension test under assigned lighting. &
--- &
--- &
With illuminance near-matched ($\sim$500\,lux) and CCT in the cool-but-not-extreme range ($\sim$3800--4200\,K), Energy-classroom students scored significantly higher on the practice ACT reading test (adjusted means $\approx$53.2 vs.\ 45.5). The improvement is attributed mainly to spectral composition (more blue-to-cyan) interacting with moderate lux and cool CCT to enhance alertness and reading comprehension. \\[6pt]
 
Zhu et al.\ (2025) &
Within-subjects lab experiment with 12 young adults reading on an e-book under two CCTs (4000 and 6500\,K) and two melanopic EDI levels per CCT, at $\sim$100\,lux vertical / 300\,lux horizontal. Reaction-time task and memory game while subjective (comfort, sleepiness, fatigue) and physiological measures were recorded. &
Illuminance on the desk held at $\sim$300\,lux; the study does not test different lux levels. Effects are driven by spectral changes at a fixed illuminance. &
At the same illuminance, 4000\,K produced more sensitivity to changes in melanopic EDI than 6500\,K: more physiological parameters and general fatigue ratings differed between low vs.\ high mel-EDI at 4000\,K, while effects at 6500\,K were smaller. &
Cognitive performance (reaction times and memory-game performance) did not show robust, systematic differences across mel-EDI or CCT conditions at 300\,lux; lighting mainly modulated alertness and comfort rather than learning/memory outcomes in this e-reading context. \\
 
\end{longtable}
\end{footnotesize}
 
\end{landscape}
 
\section{Sensitivity Analysis}
\label{app:sa}
 
A one-at-a-time (OAT) sensitivity analysis tests whether the six
predictions hold across plausible variations in eight key parameters.
Each parameter is swept across $N_\mathrm{pts}=7$ evenly spaced values
spanning approximately $\pm 50\%$ of its nominal value
(Table~\ref{tab:sa_params}), with all other parameters held fixed and
$N_\mathrm{MC}=5$ replicates per condition. The metric for each cell
is the mean pass rate across the seven sweep values.
 
\begin{table}[H]
\centering
\caption{Parameters swept, with nominal values and ranges.}
\label{tab:sa_params}
\medskip
\begin{tabular}{llccc}
\toprule
Parameter & Symbol & Nominal & Range & Channel \\
\midrule
Legibility width      & $\tau_\ell$              & 400~lux  & 200--700   & Epistemic \\
Peak precision        & $\gamma_1$               & 6.0      & 2--12      & Affective \\
Baseline precision    & $\gamma_0$               & 2.0      & 1--4       & Affective \\
Fatigue onset         & $t_\mathrm{onset}$       & 45~min   & 15--90     & Temporal  \\
Fatigue rate          & $k$                      & 6.0      & 3--12      & Temporal  \\
Lark tolerance        & $\kappa_\mathrm{lark}$   & $-3.0$   & $-6$--$-1$ & Affective \\
Owl tolerance         & $\kappa_\mathrm{owl}$    & $+3.5$   & $+1$--$+7$ & Affective \\
Over-arousal threshold& $a^*_\mathrm{over}$      & 0.78     & 0.60--0.95 & Affective \\
\bottomrule
\end{tabular}
\end{table}
 
\begin{figure}[H]
\centering
\includegraphics[width=\textwidth]{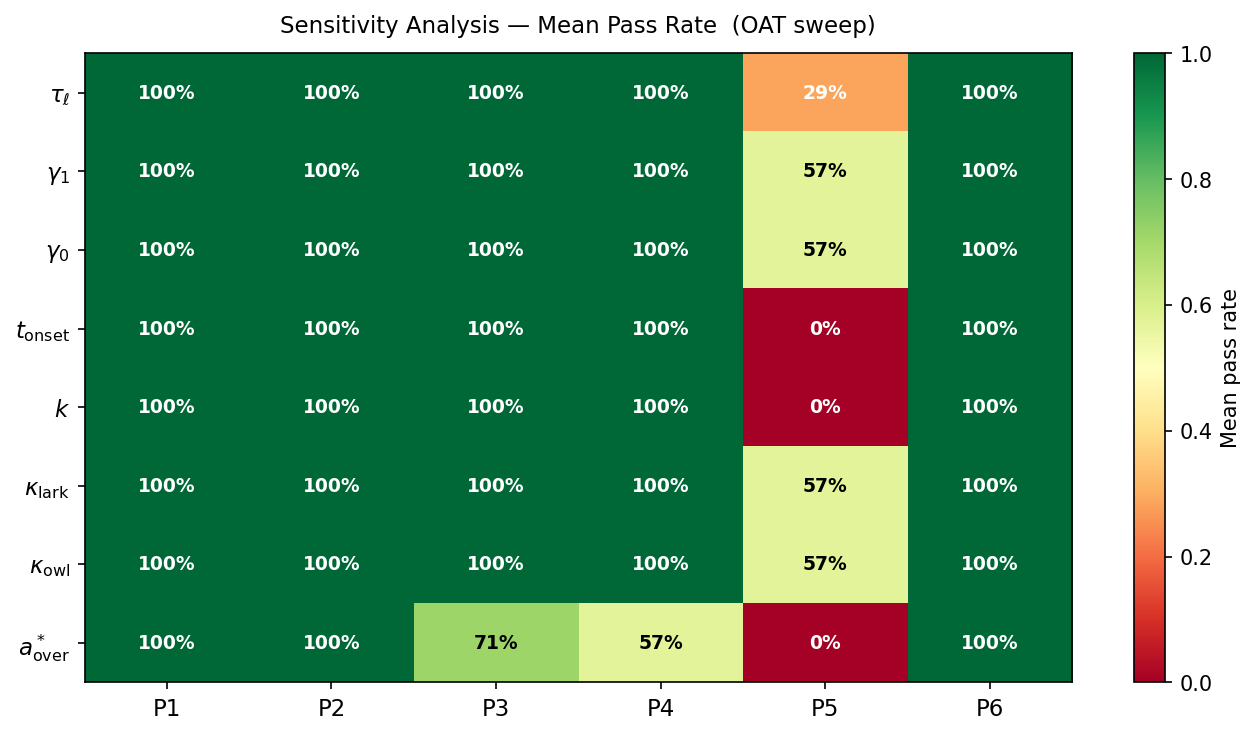}
\caption{Mean pass rate across the parameter sweep for each
         prediction. Green $\geq 80\%$; red $\leq 40\%$.}
\label{fig:sa_heatmap}
\end{figure}

\begin{figure}[H]
\centering
\includegraphics[width=\textwidth]{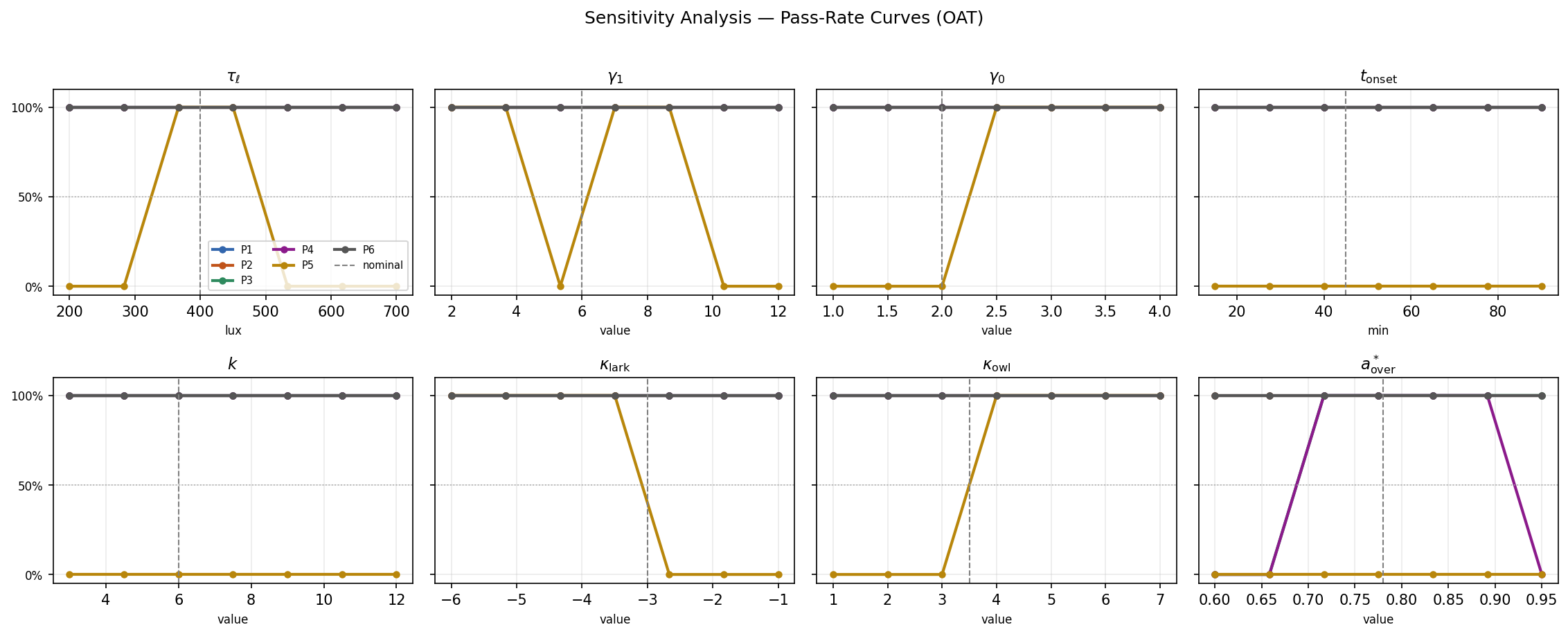}
\caption{Pass-rate curves along each parameter sweep. Dashed
         vertical lines mark the nominal value. Flat curves indicate
         structural robustness; steep drops indicate parameter
         sensitivity.}
\label{fig:sa_curves}
\end{figure}
 
Table~\ref{tab:sa_results} reports the mean pass rate for each
parameter-prediction combination.
 
\begin{table}[H]
\centering
\caption{Mean pass rate (\%) across the sweep.
         Bold $\geq 80\%$; italic $\leq 40\%$.}
\label{tab:sa_results}
\medskip
\begin{tabular}{lcccccc}
\toprule
Parameter & P1 & P2 & P3 & P4 & P5 & P6 \\
\midrule
$\tau_\ell$           & \textbf{100} & \textbf{100} & \textbf{100} & \textbf{100} & \textit{29} & \textbf{100} \\
$\gamma_1$            & \textbf{100} & \textbf{100} & \textbf{100} & \textbf{100} & 57          & \textbf{100} \\
$\gamma_0$            & \textbf{100} & \textbf{100} & \textbf{100} & \textbf{100} & 57          & \textbf{100} \\
$t_\mathrm{onset}$    & \textbf{100} & \textbf{100} & \textbf{100} & \textbf{100} & \textit{0}  & \textbf{100} \\
$k$                   & \textbf{100} & \textbf{100} & \textbf{100} & \textbf{100} & \textit{0}  & \textbf{100} \\
$\kappa_\mathrm{lark}$& \textbf{100} & \textbf{100} & \textbf{100} & \textbf{100} & 57          & \textbf{100} \\
$\kappa_\mathrm{owl}$ & \textbf{100} & \textbf{100} & \textbf{100} & \textbf{100} & 57          & \textbf{100} \\
$a^*_\mathrm{over}$   & \textbf{100} & \textbf{100} & 71           & 57           & \textit{0}  & \textbf{100} \\
\bottomrule
\end{tabular}
\end{table}
 
Three robustness tiers emerge.
 
\textbf{Tier 1 --- Structural invariants (P1, P2, P6).}
All three pass at 100\% across every sweep. P1 and P2 test properties
that follow by construction from the architecture: $\mathbf{A}$ is
chronotype-blind (P1), and $\kappa_\mathrm{lark}$ and
$\kappa_\mathrm{owl}$ have opposite signs (P2). P6 is invariant
because $\kappa_\mathrm{int.}=0$ structurally eliminates scenario
sensitivity for the intermediate chronotype regardless of all other
parameter values.
 
\textbf{Tier 2 --- Conditionally robust (P3, P4).}
Both pass at 100\% under six of the eight sweeps and degrade only when
$a^*_\mathrm{over}$ is moved above 0.85. In that regime, S2 itself
begins to induce mild over-arousal drift in $\mathbf{B}$, partially
contaminating the legibility gradient that P3 depends on, and reducing
the over-arousal penalty at the centre position that P4 depends on.
This degradation is interpretable: it occurs precisely when the specific
mechanism behind each prediction is attenuated, not when the
architecture as a whole is wrong.
 
\textbf{Tier 3 --- Parameter-sensitive (P5).}
P5 passes at 0\% under three sweeps ($t_\mathrm{onset}$, $k$,
$a^*_\mathrm{over}$) and at 29--57\% under the rest. This
prediction requires a precise balance between the owl's tolerance
advantage ($\kappa_\mathrm{owl}$), its precision advantage
($\gamma_1$), and the temporal structure of the session (fatigue
determining how much of the session both channels are informative).
Varying any of these independently tips the balance. P5 should
therefore be treated as an exploratory prediction: its confirmation
would inform the calibration of the affective and temporal parameters,
but its failure in an experiment would not challenge the three-channel
architecture.
 
P1, P2, and P6 are the appropriate targets for a minimal initial
empirical study: they would survive arbitrary reparametrisation. P3
and P4 together constitute the critical experiment. P5 is informative
about parameter values.


\begin{thebibliography}{99}

\bibitem[Abbas et~al.(2024)Abbas, Okdeh, Roufayel, Kovacic, Sabatier, Fajloun, and Abi Khattar]{Abbas2024}
Abbas, S., Okdeh, N., Roufayel, R., Kovacic, H., Sabatier, J.-M., Fajloun, Z., and Abi Khattar, Z. (2024).
\newblock Neuroarchitecture: How the perception of our surroundings impacts the brain.
\newblock \textit{Biology}, \textbf{13}(4), 220.
\newblock doi: 10.3390/biology13040220.

\bibitem[Barkmann et~al.(2012)Barkmann, Wessolowski, and Schulte-Markwort]{BarkmannEtAl2012}
Barkmann, C., Wessolowski, N., and Schulte-Markwort, M. (2012).
\newblock Applicability and efficacy of variable light in schools.
\newblock \textit{Physiology \& Behavior}, \textbf{105}(3), 621--627.
\newblock doi: 10.1016/j.physbeh.2011.09.020.

\bibitem[Blume et~al.(2019)Blume, Garbazza, and Spitschan]{Blume2019}
Blume, C., Garbazza, C., and Spitschan, M. (2019).
\newblock Effects of light on human circadian rhythms, sleep and mood.
\newblock \textit{Somnologie}, \textbf{23}(3), 147--156.
\newblock doi: 10.1007/s11818-019-00215-x.

\bibitem[Boyce(2010)]{Boyce2010}
Boyce, P.~R. (2010).
\newblock Review: The impact of light in buildings on human health.
\newblock \textit{Indoor and Built Environment}, \textbf{19}(1), 8--20.
\newblock doi: 10.1177/1420326X09358028.

\bibitem[Boyce(2022)]{Boyce2022}
Boyce, P.~R. (2022).
\newblock Light, lighting and human health.
\newblock \textit{Lighting Research \& Technology}, \textbf{54}(2), 101--144.
\newblock doi: 10.1177/14771535211010267.

\bibitem[Campbell et~al.(2023)Campbell, Navarrete, and Hattar]{CampbellEtAl2023}
Campbell, I., Navarrete, J., and Hattar, S. (2023).
\newblock Light as a modulator of non-image-forming brain functions: Positive and negative impacts of increasing light availability.
\newblock \textit{Frontiers in Neural Circuits}, \textbf{17}, 1105037.
\newblock doi: 10.3389/fncir.2023.1105037.

\bibitem[Campbell et~al.(2024)Campbell et~al.]{CampbellEtAl2024}
Campbell, I. et~al. (2024).
\newblock Regional response to light illuminance across the human subcortical visual system inferred from 7T fMRI.
\newblock \textit{Proceedings of the National Academy of Sciences of the United States of America}, \textbf{121}(43), e2406465121.
\newblock doi: 10.1073/pnas.2406465121.

\bibitem[Castilla et~al.(2023)Castilla, Higuera-Trujillo, and Llinares]{CastillaEtAl2023}
Castilla, N., Higuera-Trujillo, J.~L., and Llinares, C. (2023).
\newblock The effects of illuminance on students' memory: A neuroarchitecture study.
\newblock \textit{Building and Environment}, \textbf{228}, 109833.

\bibitem[Chauca et~al.(2024)Chauca, Mendoza, Moyano, Piedra, Vega, and S\'anchez]{ChaucaEtAl2024}
Chauca, M., Mendoza, E., Moyano, O., Piedra, L., Vega, M., and S\'anchez, A. (2024).
\newblock Improvement of student performance based on the lighting conditions of learning spaces: A systematic review analysis.
\newblock \textit{Journal of Infrastructure, Policy and Development}, \textbf{8}(16), 10619.
\newblock doi: 10.24294/jipd10619.

\bibitem[Chellappa et~al.(2011)Chellappa, Steiner, Oelhafen, Lang, Götz, Krebs, and Cajochen]{ChellappaEtAl2011}
Chellappa, S.~L., Steiner, R., Oelhafen, P., Lang, D., Götz, T., Krebs, J., and Cajochen, C. (2011).
\newblock Non-visual effects of light on melatonin, alertness and cognitive performance: Can blue-enriched light keep us alert?
\newblock \textit{PLoS ONE}, \textbf{6}(1), e16429.
\newblock doi: 10.1371/journal.pone.0016429.

\bibitem[Choi and Suk(2016)]{ChoiSuk2016}
Choi, K. and Suk, H.-J. (2016).
\newblock Dynamic lighting system for the learning environment: Performance of elementary students.
\newblock \textit{Optics Express}, \textbf{24}(10).
\newblock doi: 10.1364/OE.24.00A907.

\bibitem[{\v C}upkov\'a et~al.(2019){\v C}upkov\'a, Kaj\'ati, Mocnej, Papcun, Koziorek, and Zolotov\'a]{Cupkova2019}
{\v C}upkov\'a, D., Kaj\'ati, E., Mocnej, J., Papcun, P., Koziorek, J., and Zolotov\'a, I. (2019).
\newblock Intelligent human-centric lighting for mental wellbeing improvement.
\newblock \textit{International Journal of Distributed Sensor Networks}, \textbf{15}(9), 1550147719875878.
\newblock doi: 10.1177/1550147719875878.

\bibitem[Dijk and Archer(2009)]{DijkArcher2009}
Dijk, D.-J. and Archer, S.~N. (2009).
\newblock Light, sleep, and circadian rhythms: Together again.
\newblock \textit{PLoS Biology}, \textbf{7}(6), e1000145.
\newblock doi: 10.1371/journal.pbio.1000145.

\bibitem[Djebbara et~al.(2019)Djebbara, Fich, Petrini, and Gramann]{Djebbara2019}
Djebbara, Z., Fich, L.~B., Petrini, L., and Gramann, K. (2019).
\newblock Sensorimotor brain dynamics reflect architectural affordances.
\newblock \textit{Proceedings of the National Academy of Sciences of the United States of America}, \textbf{116}(29), 14769--14778.
\newblock doi: 10.1073/pnas.1900648116.

\bibitem[Djebbara et~al.(2021)Djebbara, Fich, and Gramann]{Djebbara2021}
Djebbara, Z., Fich, L.~B., and Gramann, K. (2021).
\newblock The brain dynamics of architectural affordances during transition.
\newblock \textit{Scientific Reports}, \textbf{11}, 2796.
\newblock doi: 10.1038/s41598-021-82504-w.

\bibitem[Easterbrook(1959)]{Easterbrook1959}
Easterbrook, J.~A. (1959).
\newblock The effect of emotion on cue utilization and the organization of behavior.
\newblock \textit{Psychological Review}, \textbf{66}(3), 183--201.
\newblock doi: 10.1037/h0047707.

\bibitem[Eberhard(2009a)]{Eberhard2009a}
Eberhard, J.~P. (2009a).
\newblock \textit{Brain Landscape: The Coexistence of Neuroscience and Architecture}.
\newblock Oxford University Press, Oxford.

\bibitem[Eberhard(2009b)]{Eberhard2009b}
Eberhard, J.~P. (2009b).
\newblock Applying neuroscience to architecture.
\newblock \textit{Neuron}, \textbf{62}(6), 753--756.
\newblock doi: 10.1016/j.neuron.2009.06.001.

\bibitem[Elliot(2015)]{Elliot2015}
Elliot, A.~J. (2015).
\newblock Color and psychological functioning: A review of theoretical and empirical work.
\newblock \textit{Frontiers in Psychology}, \textbf{6}, 368.
\newblock doi: 10.3389/fpsyg.2015.00368.

\bibitem[Friston(2010)]{Friston2010}
Friston, K. (2010).
\newblock The free-energy principle: a rough guide to the brain?
\newblock \textit{Nature Reviews Neuroscience}, \textbf{11}(2), 127--138.
\newblock doi: 10.1038/nrn2787.

\bibitem[Friston(2013)]{Friston2013}
Friston, K. (2013).
\newblock Life as we know it.
\newblock \textit{Journal of the Royal Society Interface}, \textbf{10}(86), 20130475.
\newblock doi: 10.1098/rsif.2013.0475.

\bibitem[Friston(2019)]{Friston2019}
Friston, K. (2019).
\newblock A free energy principle for a particular physics.
\newblock \textit{arXiv preprint}, arXiv:1906.10184.
\newblock doi: 10.48550/arXiv.1906.10184.

\bibitem[Grant et~al.(2021)Grant, Kent, Mayer, Stickgold, Lockley, and Rahman]{GrantEtAl2021}
Grant, L.~K., Kent, B.~A., Mayer, M.~D., Stickgold, R., Lockley, S.~W., and Rahman, S.~A. (2021).
\newblock Daytime exposure to short wavelength-enriched light improves cognitive performance in sleep-restricted college-aged adults.
\newblock \textit{Frontiers in Neurology}, \textbf{12}, 624217.
\newblock doi: 10.3389/fneur.2021.624217.

\bibitem[Guarnieri(2024)]{Guarnieri2024}
Guarnieri, T. (2024).
\newblock Light sensing beyond vision: Focusing on a possible role for the FICZ/AhR complex in skin optotransduction.
\newblock \textit{Cells}, \textbf{13}(13), 1082.
\newblock doi: 10.3390/cells13131082.

\bibitem[Hatori et~al.(2017)Hatori, Gronfier, Van Gelder, Bernstein, Carreras, Panda, Marks, Sliney, Hunt, Hirota, Furukawa, and Tsubota]{Hatori2017}
Hatori, M., Gronfier, C., Van Gelder, R.~N., Bernstein, P.~S., Carreras, J., Panda, S., Marks, F., Sliney, D., Hunt, C.~E., Hirota, T., Furukawa, T., and Tsubota, K. (2017).
\newblock Global rise of potential health hazards caused by blue light-induced circadian disruption in modern aging societies.
\newblock \textit{npj Aging and Mechanisms of Disease}, \textbf{3}, 9.
\newblock doi: 10.1038/s41514-017-0010-2.

\bibitem[Huiberts et~al.(2015)Huiberts, Smolders, and de~Kort]{HuibertsEtAl2015}
Huiberts, L.~M., Smolders, K.~C.~H.~J., and de~Kort, Y.~A.~W. (2015).
\newblock Shining light on memory: Effects of bright light on working memory performance.
\newblock \textit{Behavioural Brain Research}, \textbf{294}, 234--245.
\newblock doi: 10.1016/j.bbr.2015.07.045.

\bibitem[Jung et~al.(2024)Jung, An, and Hong]{JungEtAl2024}
Jung, D., An, J., and Hong, T. (2024).
\newblock Exploring the relationship between office lighting, cognitive performance, and psychophysiological responses: A multidimensional approach.
\newblock \textit{Building and Environment}, \textbf{263}, 111863.

\bibitem[Keis et~al.(2014)Keis, Helbig, Streb, and Hille]{KeisEtAl2014}
Keis, O., Helbig, H., Streb, J., and Hille, K. (2014).
\newblock Influence of blue-enriched classroom lighting on students' cognitive performance.
\newblock \textit{Trends in Neuroscience and Education}, \textbf{3}(3--4), 86--92.
\newblock doi: 10.1016/j.tine.2014.09.001.

\bibitem[Kızıltunalı(2023)]{Kiziltunali2023}
Kızıltunalı, B. (2023).
\newblock A literature review: The impact of light on students' learning performance.
\newblock \textit{Humanising Language Teaching Magazine}, August 2023.

\bibitem[Knez(1995)]{Knez1995}
Knez, I. (1995).
\newblock Effects of indoor lighting on mood and cognition.
\newblock \textit{Journal of Environmental Psychology}, \textbf{15}(1), 39--51.
\newblock doi: 10.1016/0272-4944(95)90013-6.

\bibitem[Kumari et~al.(2023)Kumari, Das, Babaei, Rokni, and Goldust]{Kumari2023}
Kumari, J., Das, K., Babaei, M., Rokni, G.~R., and Goldust, M. (2023).
\newblock The impact of blue light and digital screens on the skin.
\newblock \textit{Journal of Cosmetic Dermatology}, \textbf{22}(4), 1185--1190.
\newblock doi: 10.1111/jocd.15576.

\bibitem[Lasauskaite et~al.(2025)Lasauskaite, Wüst, Schöllhorn, Richter, and Cajochen]{LasauskaiteEtAl2025}
Lasauskaite, R., Wüst, L.~N., Schöllhorn, I., Richter, M., and Cajochen, C. (2025).
\newblock Non-image-forming effects of daytime electric light exposure in humans: A systematic review and meta-analyses of physiological, cognitive, and subjective outcomes.
\newblock \textit{LEUKOS}.
\newblock doi: 10.1080/15502724.2025.2493669.

\bibitem[LeGates et~al.(2014)LeGates, Fernandez, and Hattar]{LeGates2014}
LeGates, T.~A., Fernandez, D.~C., and Hattar, S. (2014).
\newblock Light as a central modulator of circadian rhythms, sleep and affect.
\newblock \textit{Nature Reviews Neuroscience}, \textbf{15}(7), 443--454.
\newblock doi: 10.1038/nrn3743.

\bibitem[Li and Yao(2025)]{LiYao2025}
Li, Z. and Yao, R. (2025).
\newblock Assessing the impact of office artificial lighting on young adults' reading performance through eye-tracking analysis.
\newblock \textit{Building and Environment}, \textbf{285}, 113532.

\bibitem[Livingston(2022)]{Livingston2022}
Livingston, J. (2022).
\newblock \textit{Designing with Light: The Art, Science, and Practice of Architectural Lighting Design}.
\newblock 2nd edn.
\newblock Wiley, Hoboken, NJ.

\bibitem[Mott et~al.(2012)Mott, Robinson, Rutherford, and Burnette]{MottEtAl2012}
Mott, M., Robinson, D.~H., Rutherford, A., and Burnette, J. (2012).
\newblock Illuminating the effects of dynamic lighting on student learning.
\newblock \textit{SAGE Open}, April--June, 1--9.

\bibitem[Namjoshi(2026)]{namjoshi2026fundamentals}
Namjoshi, S.~V. (2026).
\textit{Fundamentals of Active Inference: Principles, Algorithms,
and Applications of the Free Energy Principle for Engineers}.
The MIT Press.
ISBN: 9780262050951.

\bibitem[Naylor and Firth(2008)]{NaylorFirth2008}
Naylor, K. and Firth, A.~Y. (2008).
\newblock The effect of coloured paper and lighting on the rate of reading in an adult student population.
\newblock \textit{British and Irish Orthoptic Journal}, \textbf{5}, 54--57.

\bibitem[Nol\'e Fajardo et~al.(2023)Nol\'e Fajardo, Higuera-Trujillo, and Llinares]{NoleFajardoEtAl2023}
Nol\'e Fajardo, M.~L., Higuera-Trujillo, J.~L., and Llinares, C. (2023).
\newblock Lighting, colour and geometry: Which has the greatest influence on students' cognitive processes?
\newblock \textit{Frontiers of Architectural Research}, \textbf{12}, 575--586.
\newblock doi: 10.1016/j.foar.2023.02.003.

\bibitem[Park et~al.(2013)Park, Ha, Ryu, Kim, Jung, and Kim]{ParkEtAl2013}
Park, J.~Y., Ha, R.-Y., Ryu, V., Kim, E., Jung, Y.-C., and Kim, S.-Y. (2013).
\newblock Effects of color temperature and brightness on electroencephalogram alpha activity in a polychromatic light-emitting diode.
\newblock \textit{Clinical Psychopharmacology and Neuroscience}, \textbf{11}(3), 126--131.
\newblock doi: 10.9758/cpn.2013.11.3.126.

\bibitem[Parr et al.(2022)]{parr2022active}
Parr, T., Pezzulo, G., \& Friston, K.~J. (2022).
\textit{Active Inference: The Free Energy Principle in Mind, Brain,
and Behavior}.
The MIT Press.
ISBN: 9780262045353.

\bibitem[Plummer(2016)]{Plummer2016}
Plummer, H. (2016).
\newblock \textit{The Experience of Architecture}.
\newblock Thames \& Hudson, London.

\bibitem[Possati(2026)]{possati2026design}
Possati, L.~M. (2026).
\textit{Design for Entropy: Active Inference and Technology}.
The MIT Press.
ISBN: 9780262056267.

\bibitem[Ram and Bhardwaj(2018)]{RamBhardwaj2018}
Ram, M.~S. and Bhardwaj, R. (2018).
\newblock Effect of different illumination sources on reading and visual performance.
\newblock \textit{Journal of Ophthalmic and Vision Research}, \textbf{13}(1), 44--49.
\newblock doi: 10.4103/jovr.jovr\_50\_17.

\bibitem[Pallasmaa(2015)]{Pallasmaa2015}
Robinson, S. and Pallasmaa, J., eds. (2015).
\newblock \textit{Mind in Architecture: Neuroscience, Embodiment, and the Future of Design}.
\newblock MIT Press, Cambridge, MA.

\bibitem[Ru et~al.(2021)Ru, Smolders, Chen, Zhou, and de~Kort]{RuEtAl2021}
Ru, T., Smolders, K.~C.~H.~J., Chen, Q., Zhou, G., and de~Kort, Y.~A.~W. (2021).
\newblock Diurnal effects of illuminance on performance: Exploring the moderating role of cognitive domain and task difficulty.
\newblock \textit{Lighting Research \& Technology}, \textbf{53}, 727--747.
\newblock doi: 10.1177/1477153521990645.

\bibitem[Shi et~al.(2025)Shi, Hu, and Xue]{ShiEtAl2025}
Shi, Y., Hu, J., and Xue, T. (2025).
\newblock Light, opsins, and life: Mammalian photophysiological functions beyond image perception.
\newblock \textit{Neuron}, \textbf{113}(19), 3108--3128.
\newblock doi: 10.1016/j.neuron.2025.05.025.

\bibitem[Shishegar and Boubekri(2022)]{ShishegarBoubekri2022}
Shishegar, N. and Boubekri, M. (2022).
\newblock Lighting up living spaces to improve mood and cognitive performance in older adults.
\newblock \textit{Journal of Environmental Psychology}, \textbf{82}, 101845.
\newblock doi: 10.1016/j.jenvp.2022.101845.

\bibitem[Sholanke et~al.(2021)Sholanke, Fadesere, and Elendu]{Sholanke2021}
Sholanke, A.~B., Fadesere, O., and Elendu, D. (2021).
\newblock The role of artificial lighting in architectural design: A literature review.
\newblock \textit{IOP Conference Series: Earth and Environmental Science}, \textbf{665}(1), 012008.
\newblock doi: 10.1088/1755-1315/665/1/012008.

\bibitem[T\"ahk\"am\"o et~al.(2019)T\"ahk\"am\"o, Partonen, and Pesonen]{Tahkamo2019}
T\"ahk\"am\"o, L., Partonen, T., and Pesonen, A.-K. (2019).
\newblock Systematic review of light exposure impact on human circadian rhythm.
\newblock \textit{Chronobiology International}, \textbf{36}(2), 151--170.
\newblock doi: 10.1080/07420528.2018.1527773.

\bibitem[Vandewalle et~al.(2009)Vandewalle, Maquet, and Dijk]{VandewalleEtAl2009}
Vandewalle, G., Maquet, P., and Dijk, D.-J. (2009).
\newblock Light as a modulator of cognitive brain function.
\newblock \textit{Trends in Cognitive Sciences}, \textbf{13}(10), 429--438.
\newblock doi: 10.1016/j.tics.2009.07.004.

\bibitem[Vries de (2026)]{devries2026active}
de Vries, B. (2026).
Active inference for physical {AI} agents: An engineering perspective.
\textit{arXiv preprint arXiv:2603.20927}.

\bibitem[Wang et~al.(2022)Wang, Sanches de Oliveira, Djebbara, and Gramann]{Wang2022}
Wang, S., Sanches de Oliveira, G., Djebbara, Z., and Gramann, K. (2022).
\newblock The embodiment of architectural experience: A methodological perspective on neuro-architecture.
\newblock \textit{Frontiers in Human Neuroscience}, \textbf{16}, 833528.
\newblock doi: 10.3389/fnhum.2022.833528.

\bibitem[Yan et~al.(2019)Yan, Lonstein, and Nunez]{Yan2019}
Yan, L., Lonstein, J.~S., and Nunez, A.~A. (2019).
\newblock Light as a modulator of emotion and cognition: Lessons learned from studying a diurnal rodent.
\newblock \textit{Hormones and Behavior}, \textbf{111}, 78--86.
\newblock doi: 10.1016/j.yhbeh.2018.09.003.

\bibitem[Yerkes and Dodson(1908)]{Yerkes1908}
Yerkes, R.~M., and Dodson, J. (1908).
\newblock The relation of strength of stimulus to rapidity of habit formation.
\newblock {\it The Journal of Comparative Neurology and Psychology}, 18, 459--482.
\newblock doi:10.1002/cne.920180503

\bibitem[Zhou and Pan(2023)]{ZhouPan2023}
Zhou, A. and Pan, Y. (2023).
\newblock Effects of indoor lighting environments on paper reading efficiency and brain fatigue: An experimental study.
\newblock \textit{Frontiers in Built Environment}, \textbf{9}, 1303028.
\newblock doi: 10.3389/fbuil.2023.1303028.

\bibitem[Zhu et~al.(2019)Zhu, Yang, Yao, Xiong, Li, Zhou, and Ma]{ZhuEtAl2019}
Zhu, Y., Yang, M., Yao, Y., Xiong, X., Li, X., Zhou, G., and Ma, N. (2019).
\newblock Effects of illuminance and correlated color temperature on daytime cognitive performance, subjective mood, and alertness in healthy adults.
\newblock \textit{Environment and Behavior}, \textbf{51}(2), 199--230.
\newblock doi: 10.1177/0013916517738077.

\bibitem[Zhu et~al.(2025)Zhu, Tu, Wang, Zhu, and Shi]{ZhuEtAl2025}
Zhu, N., Tu, Y., Wang, L., Zhu, X., and Shi, Y. (2025).
\newblock The effect of melanopic illuminance of illuminated environment on visual comfort, alertness and cognitive performance during e-book reading.
\newblock \textit{Lighting Research \& Technology}, \textbf{XX}, 1--27.
\newblock doi: 10.1177/14771535251351892.

\end{thebibliography}
\end{document}